\documentclass[preprint]{aastex}
\received{}
\revised{}
\accepted{}
\journalid{}{}
\articleid{}{}
\paperid{}
\cpright{}{}
\ccc{}
\lefthead{Matthews \& de Grijs}
\righthead{UGC~10043}
\begin{document}

\newcommand{\ang}{\rm \AA}
\newcommand{\msun}{\mbox{${\cal M}_\odot$}}
\newcommand{\lsun}{\mbox{${\cal L}_\odot$}}
\newcommand{\kms}{\mbox{km s$^{-1}$}}
\newcommand{\HI}{\mbox{H\,{\sc i}}}
\newcommand{\HA}{H$\alpha$}
\newcommand{\mhi}{\mbox{$M_{\rm HI}$}}
\newcommand{\HII}{\mbox{H\,{\sc ii}}}
\newcommand{\CaII}{\mbox{Ca\,{\sc ii}}}
\newcommand{\MgII}{\mbox{Mg\,{\sc ii}}}
\newcommand{\NII}{\mbox{N\,{\sc ii}}}
\newcommand{\CIV}{\mbox{C\,{\sc iv}}}
\newcommand{\am}[2]{$#1'\,\hspace{-1.7mm}.\hspace{.0mm}#2$}
\newcommand{\as}[2]{$#1''\,\hspace{-1.7mm}.\hspace{.0mm}#2$}
\newcommand{\ad}[2]{$#1^{\circ}\,\hspace{-1.7mm}.\hspace{.0mm}#2$}
\newcommand{\lsim}{~\rlap{$<$}{\lower 1.0ex\hbox{$\sim$}}}
\newcommand{\gsim}{~\rlap{$>$}{\lower 1.0ex\hbox{$\sim$}}}
\newcommand{\dark}{$M_{HI}/L_{B}$}
\newcommand{\Sbc}{Sbc}

\title{Optical Imaging and Spectroscopy of the Edge-On 
\Sbc\ Galaxy UGC~10043:
Evidence for a Galactic Wind and a Peculiar Triaxial Bulge}

\vskip0.5cm
\author{L. D. Matthews\altaffilmark{1,2}}
\author{R. de Grijs\altaffilmark{3}}

\altaffiltext{1}{Harvard-Smithsonian Center for Astrophysics, 60
Garden Street, MS-42, Cambridge, MA 02138 USA,
Electronic mail: lmatthew@cfa.harvard.edu}
\altaffiltext{2}{Visiting Astronomer, Kitt Peak National Observatory.
National Optical Observatories is operated by the Association of
Universities for Research in Astronomy (AURA), Inc., under a
cooperative agreement with the National Science Foundation.}
\altaffiltext{3}{Department of Physics \& Astronomy, the University of
Sheffield, Hicks Building, Hounsfield Road, Sheffield S3~7RH, United
Kingdom,
Electronic mail: R.DeGrijs@sheffield.ac.uk}

\singlespace
\tighten
\begin{abstract}
We present new optical imaging and spectroscopy of the peculiar,
edge-on Sbc galaxy UGC~10043. Using the WIYN telescope, we have
obtained $B$, $R$, and \HA+[\NII] 
images, together with DensePak integral field
spectroscopic measurements of the stellar \CaII\
infrared triplet and the H$\alpha$ and
[\NII] lines from the ionized gas.  
The imaging observations show that the inner bulge of UGC~10043
($a\le$\as{7}{5}) is elongated perpendicular to the galaxy major
axis. At larger $r$, the bulge isophotes twist to become oblate
and nearly circular, suggesting the bulge is triaxial. 
The bulge shows no clear evidence for
rotation about either its major or minor axis. The inner,
southwestern quadrant of the bulge is girdled by a narrow dust lane 
parallel to the minor axis; unsharp masking reveals that
this minor axis dust lane may be part of
an inner polar ring,
although we find no unambiguous kinematic 
evidence of orthogonally rotating material. 
The stellar disk of UGC~10043 has a rather
low optical surface brightness [$\mu(0)_{R,i}\sim$23.2 mag
arcsec$^{-2}$], a small scale
height ($h_{z}=395$~pc for $D=33.4$~Mpc), and a mild integral sign warp. 
A dusty, inner disk
component  that appears tilted relative to 
the outlying disk is also seen.
The \HA\ and [\NII] emission lines in UGC~10043 resolve into multiple
velocity components, indicating the presence 
of a large-scale galactic wind with an outflow
velocity of $V_{\rm out}\gsim$104~\kms. \HA+[\NII] imaging reaffirms this
picture by revealing ionized gas extended to $|z|\sim$3.5 kpc in the
form of a  roughly biconical structure. The [\NII]/\HA\ line intensity ratio
increases with increasing distance from the plane, reaching values as
high as 1.7. Unlike most galaxies with large-scale winds, 
UGC~10043 has only a modest global star formation rate
($\lsim1M_{\odot}$ yr$^{-1}$), implying the wind is powered by a
rather feeble central starburst.
We discuss evolutionary scenarios that could
account for both the structural complexities of UGC~10043 and its
large-scale wind. The most plausible scenarios require a
major accretion or merger event at least a few Gyr ago.

\end{abstract}


\section{Background\protect\label{intro}}
UGC~10043 is a nearby galaxy ($D\approx$33.4~Mpc; see Table~1)
that has been classified as an Sbc spiral
seen almost directly edge-on
(e.g., de Grijs \& van der Kruit 1996).  
However, recent high-resolution {\it Hubble Space Telescope} ($HST$)
images obtained through $F300W$ and $F814W$ filters
(mid-ultraviolet and $I$-equivalent bands) by  Windhorst et al. (2002) 
highlighted 
some morphological peculiarities of this object,
including an extremely
thin disk that
appears very dim in the mid-ultraviolet, a bulge with prolate isophotes,
and a  narrow dust lane 
running parallel to the minor axis. 

Vertical elongation is not a common feature of spiral bulges (e.g.,
Bertola, Vietri, \& Zeilinger 1991). Indeed, most bulges are significantly
flattened by rotation, and many appear ``disky'', with 
scale heights only slightly larger
than their host disks (e.g., Kormendy 1993). A notable 
exception is the case of certain
early-type spirals where ``orthogonally decoupled bulges'' have been
recently discovered (Bertola et al. 1999;
Sarzi et al. 2000; Corsini et
al. 2003; see also Arnaboldi et al. 1993; 
Reshetnikov, Fa\'undez-Abans, \& de Oliveira-Abans
2002). Another class of galaxies displaying vertically 
elongated inner components are the 
polar ring galaxies (e.g., Whitmore et al. 1990), 
although in this latter case, 
the ``prolate bulge'' is
actually an S0 or similar galaxy,
whose rotation is orthogonal to an orbiting ring or disk of
stars and gas (see the review by Sparke 2002).  

A common trait of both
the orthogonally decoupled bulge systems and the polar ring galaxies 
is that both classes of objects contain a structural component
whose angular momentum vector is roughly parallel to the
major axis of the host galaxy. This
in turn suggests these systems could not have formed from 
the collapse of a single
protogalactic cloud, but instead have undergone a 
``second event'' after a precursor of the system was significantly evolved
(e.g., Schweizer, Whitmore, \& Rubin 1983; Bertola et al. 1999;
Bournaud \& Combes 2003). 
Another group of
objects belonging to the class of ``multi-spin
galaxies'' are the minor-axis dust lane ellipticals. These are observed
to have modest rings of gas and dust orbiting about their short axis
(e.g., Hawarden et al. 1981; Bertola 1987; Oosterloo et al. 2002).
The peculiar minor axis dust lane of UGC~10043 causes its bulge to
bear an interesting 
resemblance to these latter systems, while its elongated 
bulge and massive stellar disk suggest a more extreme version of these
systems, and a possible link
to the polar ring or
orthogonally-decoupled bulge galaxies.

Polar rings, orthogonally decoupled bulges, and minor axis dust lanes
are all relatively rare, but these systems are nonetheless of
considerable interest from a galaxy evolution perspective. 
For example, the ability to measure the
kinematics of a galaxy along two orthogonal axes permits unique constraints on
the shape of the potential and of the dark matter halo (e.g., Tohline,
Simonson, \& Caldwell 1982; Sackett \& Sparke 1990; Sparke 1996; Reshetnikov \&
Sotnikova 1997). Multi-spin galaxies
serve as important evidence that there is no single recipe 
for the formation of
a disk+spheroid system, and demonstrate that such galaxies are likely to  
be built  and modified by
a variety of processes---all of which need to be accounted for in a
complete picture of galaxy formation. In addition, we now have significant
observational evidence that ``second events'' have played a key role in the
evolution of nearly all massive galaxies (e.g.,
Freeman \& Bland-Hawthorn 2002 and references therein).
However, because the
conditions required to form a stable, multi-spin
galaxy from an accretion or merger event 
are rather restricted, these systems provide particularly
important laboratories for exploring how
initial conditions and the properties of the participating galaxies
influence the outcome of these episodes. 
Studies of multi-spin
galaxies may therefore 
provide important clues on the roles
accretion and minor mergers played in shaping a much wider array of
observed galaxy types
(e.g., Bettoni \& Galletta 1991).

The morphological similarities between UGC~10043 and certain
multi-spin galaxies strongly suggest a relationship to these objects, and
imply that UGC~10043 has had a 
rather different formation history from ``prototypical'' 
Sbc galaxies such as the Milky Way or NGC~891. 
To investigate this possibility further, we have obtained new
optical imaging and
spectroscopic observations of UGC~10043. This paper describes the
analysis of these data, and uses them as a first step
toward building a picture of the evolutionary history of this
galaxy.

\section{Imaging Observations and Photometric Calibration\protect\label{phot}}
Some basic properties of UGC~10043 are summarized in Table~1.
The $HST$ imaging observations of Windhorst et al. (2002)
comprised only very short exposures covering 
a small field-of-view  (\am{2}{7}). Therefore we  have used the Mini-Mosaic
Camera (Mini-Mo) 
on the WIYN\footnote{The WIYN Observatory is a joint facility of the
University of Wisconsin-Madison, Indiana University, Yale University, 
and the National Optical Astronomy Observatory.} 
telescope at Kitt Peak to obtain new
imaging and photometry of UGC~10043 in order to permit a
more detailed morphological and photometric study.

The Mini-Mo images were obtained on 
2001 April 16 through 
Harris $B$ and $R$ filters, and a narrow-band (72\ang)
filter (W016), centered at $\sim$6620\ang\ and
transmitting the redshifted \HA+[\NII] emission. 
Exposure times were 750 seconds in $R$, 1000 seconds in
$B$, and 1200+800 seconds in \HA. 

Mini-Mo consists of two thinned SITe 4096$\times$2048 CCDs 
separated by a small
(\as{7}{1}) gap. The plate scale is \as{0}{141}/pixel and the field-of-view
is \am{9}{6}$\times$\am{9}{6}. Each of the CCDs is read out by
two amplifiers whose mean gain and readout noise are
1.4$e^{-}$ ADU$^{-1}$ and 5.5$e^{-}$, respectively. 

Reduction of the images was performed in 
IRAF\footnote{IRAF is distributed by the 
National Optical
Astronomy Observatories, which is operated by the Associated
Universities for Research in Astronomy, Inc., under cooperative
agreement with the National Science Foundation.}
utilizing tasks within the `mscred' package.
Flatfielding was accomplished using a median of several dome flats
obtained in the appropriate filter. In all filters,  
small gain variations led to flatfield gradients of order 2\% across
the two quadrants of the CCD containing UGC~10043. These gradients were
reduced to
$\lsim$1\% by means of  two-dimensional,
second order Legendre polynomial fits to the image.

The Mini-Mo $B$ and $R$ 
observations were acquired under dark, photometric conditions
with seeing of \as{0}{6} for the $R$ image and \as{0}{7} for the $B$
image.  For the $B$ and $R$ observations,
photometric calibration and conversion to a standard system 
was performed by observing sets of standard
stars from
Landolt (1992) at three different airmasses (1.1, 1.2, and 1.5). 
The photometric solution is summarized in Table~2.

Photometry of UGC~10043 was performed through elliptical
apertures after removing cosmic rays, foreground stars, and bad CCD
columns. The results are
presented in Table~1. The 
photometric errors were computed following Matthews \&
Gallagher (1997) and also include the scatter in the photometric 
solution.  The apparent magnitudes were corrected for Galactic
foreground extinction following Schlegel, Finkbeiner, \& Davis (1998).
The resulting apparent magnitudes  in Table~1 
agree to within errors with those
published in the Third Reference Catalogue of Bright Galaxies ($m_{B}=
14.62\pm0.20$; de Vaucouleurs et al. 1991) and by Pohlen (2001;
$m_{R}=13.53$). 

Seeing during the narrow-band (\HA+[\NII]) observations  
was $\sim$\as{0}{8}. The two narrow-band exposures were averaged, and cosmic
rays were removed by hand in the vicinity of UGC~10043.
Continuum subtraction was performed using the $R$-band image,
after convolving it with a Gaussian to match the seeing
of the narrow-band images. An \HA\ 
luminosity for UGC~10043 was  estimated
using the $R$-band calibration and the Harris~$R$ and W016 filter transmission
curves. [\NII]/\HA=0.43 was assumed
(see Section~\ref{ratios}), 
yielding $L_{{\rm H}\alpha}\approx1.7\times 10^{40}$ erg s$^{-1}\pm$20\%
(uncorrected for internal extinction or underlying stellar absorption).

\section{Imaging Results: Optical Properties and Morphology of 
UGC~10043\protect\label{optmorph}}
The WIYN $R$-band image of  UGC~10043 is
presented in Figures~\ref{fig:wiynimages}a \& \ref{fig:wiynimagewide},
while Figure~\ref{fig:wiynimages}b shows a $B$+$R$ composite image.
These images highlight with improved signal-to-noise
several of the features visible in the $HST$ images of Windhorst et
al. (2002), and also showcase several new features.

\subsection{The Disk of UGC~10043\protect\label{disk}}
While its size and mass are typical for Sbc spirals (Table~1),
a number of characteristics differentiate  UGC~10043 from most Sbc
spiral galaxies viewed edge-on. First, the disk
component of UGC~10043 is strikingly 
thin for a galaxy of this Hubble type. 
To estimate the disk scale height, we extracted
light profiles from the $R$-band image
in 15-pixel-wide strips perpendicular to the major axis
at locations 
$r$=45$''$ southeast and $r=34''$ northwest of the galaxy
center. These locations were chosen to minimize contamination from
dust clumps and bright \HII\ regions.
Fitting exponential functions to these
light profiles over the interval $z=2''-18''$, 
we derive an estimate of the global disk scale height of
$h_{z}=$\as{2}{44}$\pm$\as{0}{04} ($\approx$395~pc), 
with the values found for the two extracted strips
differing by only $\sim$3\%. After correction to the same distance
scale, this value is in good agreement with the
$I$-band determination of de Grijs \& van der Kruit (1996;
412$\pm$10~pc) but slightly smaller than the value determined in the
$R$-band by Pohlen (2001; 453~pc).
Such a global disk scale height 
is roughly a factor of two smaller than
typical global values measured for the disks of 
Sb-Sbc spirals (e.g., Wainscoat et al. 1989; de Grijs 1998; 
Pohlen et al. 2000), but is comparable to the old thin disk scale height
of the Milky Way (Chen et al. 2001; Siegel et al. 2002) and of many Sd
spirals (e.g., de Grijs 1998).

Appropriate corrections for internal extinction in UGC~10043 are
rather uncertain. The UGC~10043 disk is clearly dusty,
although outside the inner $\sim$30$''$,  the dust appears to be highly
clumped, and does not form a continuous dust lane; hence the amount of
dust may vary considerably with radius (see also
Section~\ref{cmap}). 
Furthermore, a significant
fraction of the galaxy light comes from the bulge component (see below) where
the dust content may be significantly different. 
Using the prescription of Tully
et al. (1998), whereby the internal extinction of a galaxy 
is estimated based on its rotational velocity, we derive total internal
extinctions of $A_{B,i}$=1.47 and $A_{R,i}$=1.02 mags, assuming $H_{0}$=70
\kms\ Mpc$^{-1}$,
$W^{i}_{R}\approx2V_{rot}\approx320$~\kms\ 
(see Table~1 and Section~\ref{radio}), and
$a/b$=7.5 (the $R$-band disk axial ratio at the observed 25 mag arcsec$^{-2}$
isophote). 
From this we estimate for
UGC~10043 total absolute magnitudes of $M_{B}\approx-19.4$ and
$M_{R}\approx-20.2$, and a $B$-band luminosity
$L_{B}\approx 9.0 \times10^{9}~L_{\odot}$.

It has been found for galaxies of a given dynamical mass that the
thinnest disks tend to be of the lowest surface brightness (e.g., Gerritsen
\& de Blok 1999; Bizyaev \&
Mitronova 2002). Consistent with this, the thin  disk of UGC~10043 is
of rather modest brightness in the $B$ and $R$ bands despite its
edge-on geometry
(Figure~\ref{fig:Rmajor}). This cannot be attributed
solely to dust extinction, as the disk also appears very faint in the
$I$-band (see Figures~3.16 \& 4.16 of Windhorst et al. 2002). By
extrapolating the disk light profile to small $r$, we estimate a 
central disk surface brightness $\mu_{R}(0)\approx$21.2 mag
arcsec$^{-2}$, consistent with the estimate of Pohlen (2001). 
Deprojected to face-on (assuming $i=90^{\circ}$), 
and corrected for internal
extinction (assuming a simple foreground screen), 
this corresponds to $\mu(0)_{R,i}\sim$23.2 mag
arcsec$^{-2}$. This value is characteristic of very late-type, low
surface brightness spirals, but is $\sim$1.5-3.0 magnitudes fainter than values
typical for ``normal'' 
Sb-Sc galaxies (e.g., de Jong 1996a; Tully \& Verheijen 1997).

Finally, we draw attention to 
one additional feature of the UGC~10043 disk:
the slight ``integral sign'' warp of its
outer disk (Figure~\ref{fig:wiynimagewide}). While
stellar warps are rather common in late-type disks, 
they become increasingly rare in galaxies with
significant bulge components (Pitesky 1991; Reshetnikov \&
Combes 1998).

\subsection{The Bulge of UGC~10043: Evidence of Triaxiality, 
Geometric Decoupling, and a Possible Inner Polar Ring\protect\label{bulge}}
Even a casual inspection of UGC~10043 quickly reveals that its disk 
is not its only intriguing feature;
its bright bulge is also unusual in that its inner isophotes
appear distinctly prolate (Figure~\ref{fig:wiynimages}a \& b), while
the  outer isophotes are nearly circular 
(Figure~\ref{fig:wiynimagewide}).
The high surface brightness of the bulge compared with the adjacent
disk regions, together with its unflattened shape and comparatively large
vertical extent, lend the impression of a structurally
distinct entity, as opposed to a 
dynamically hotter extension of the disk.
Furthermore, on the southwest side of the midplane, a distinct dust 
structure is visible perpendicular to the disk, and 
bisecting the bulge along the direction of the apparent 
minor axis. As noted in the Introduction,
this dust structure is reminiscent of those seen 
in minor axis dust lane elliptical galaxies with orthogonally rotating
gas (e.g., Bertola 1987), and thus
hints at the presence of 
material with misaligned angular momentum 
(although this has yet to be confirmed kinematically; see
Section~\ref{spectroscopy}).

\subsubsection{Non-Parametric Bulge Decomposition\protect\label{bulgedecom}}
In order to better quantify the shape of the UGC~10043 
bulge and estimate its total
luminosity, we performed a non-parametric decomposition of the $R$-band
image by fitting a series of ellipses to the bulge region.
Semi-major axes of the ellipses ranged from $a$=1$''$-20$''$. 
The regions along the midplane 
most affected by dust  were masked during the fitting
($|z|<$\as{0}{5} for $a\lsim4''$; $|z|<$\as{1}{5}-\as{2}{0} for
$a\gsim4''$, plus a $\sim$\as{0}{5}-wide swath along the minor-axis
dust lane). The
position angle, ellipticity, and center of each ellipse were allowed
to vary freely, except for $a<4''$, 
where the center was kept fixed. For $a<2''$ the fits
are rather uncertain owing to significant contamination from dust
and disk light. 

Using a two-dimensional model generated
from the ellipse fits, 
we estimate the total $R$ magnitude of the bulge component of UGC~10043 to be
$M_{R}\approx -18.5$ (assuming dust extinction within the
bulge itself is negligible). After correcting the residual disk light for internal
extinction, this yields a bulge-to-disk ratio B/D$\sim$0.3. 

Further details from the bulge fitting 
are summarized in Figure~\ref{fig:bulgefits}. 
The fits indicate that within the inner 15$''$ ($a\le$\as{7}{5}), 
the UGC~10043 bulge isophotes have a prolate
shape, with ellipticity $\epsilon = 1 - (b/a)$ decreasing 
with increasing $a$, from $\epsilon\approx0.60$ for $a<4''$ to
$\epsilon$=0.11 at $a$=\as{7}{5}
(Figure~\ref{fig:bulgefits}). Beyond $a=8''$, the
next fitted isophote underwent an abrupt position angle shift 
of 98$^{\circ}$; the position angles of the
successive outlying isophotes then became closely aligned with the disk
position angle of \ad{151}{5}$\pm$\ad{0}{5}, and 
were oblate and roughly circular
($\epsilon$=0.02-0.24, with $\epsilon$ increasing systematically
with $a$). 
These isophote shifts  suggest the bulge of UGC~10043 is inherently triaxial
(see Mihalas \& Binney 1981; Bertola, Vietri, \& Zeilinger 1991).

In the lower panel of Figure~\ref{fig:bulgefits}, we plot as a function
of ellipse semi-major axis, the dimensionless parameter ${\rm
A}_{4}/a$, where A$_{4}$ is a Fourier coefficient that characterizes
the deviation of the fitted ellipses from a true elliptical shape
(see Bender, D\"obereiner, \& M\"ollenhoff 1988). Following the
convention of Bender et
al. (1988),
negative ${\rm A}_{4}/a$ values indicate ``boxy'' (rectangular-shaped)
isophotes, while
positive values indicate ``disky'' (leaf-shaped)
isophotes. The A$_{4}$ values inside
$a<4''$ are poorly constrained and have been omitted from the
figure. In UGC~10043, a transition from boxy to disky isophotes
mirrors the decrease in ellipticity observed over the interval
$4''<a<8''$. The maximum ${\rm A}_{4}/a$ value occurs just beyond
this, near $a=9''$. Beyond $a>11''$, the isophotes show a roughly
constant ${\rm A}_{4}/a$ value of 0.1\%. Note that while disky
bulge isophotes elongated {\it along} the major axis 
are quite common (e.g., Kormendy 1993), disky
isophotes elongated {\it perpendicular} to the major axis are not (e.g.,
Bertola et al. 1991), and
suggest the possible existence of an embedded structure. However,
the diskiness of the isophotes suggests that any embedded
structure could not be a bar seen with its long axis along the
line-of-sight, since such a feature would be expected
to instead appear boxy and flattened (Combes et al. 1990;
Merrifield 1996). If the UGC~10043 bulge harbors a bar, it would therefore
need to be
oriented perpendicular to our line-of-sight, and be tumbling about
its long axis---a situation that seems dynamically improbable.

\subsubsection{Unsharp Masking\protect\label{unsharp}}
Subtraction of the bulge model derived in Section~\ref{bulgedecom}
from the original image revealed hints of additional complex
structure. Therefore,
to further probe the underlying structure of the bulge,
we employed the technique of unsharp masking
(see Erwin \& Sparke 2003 and references therein). 
One type of unsharp mask can be produced simply by
subtracting a smoothed version of an image from the original.
In our case, we subtracted from the $R$-band frame a 
version of the image that had been smoothed using a Gaussian with $\sigma$=10
pixels. A portion of the
resulting mask is shown in Figure~\ref{fig:unsharp}.

Figure~\ref{fig:unsharp} unveils a rather intriguing structure:
a possible ring or annulus of material oriented roughly 
orthogonally to the main disk. 
An unsharp mask made from the $B$-band image appears nearly
identical.  As
seen in Figure~\ref{fig:unsharp}, 
the apparent ring
forms a continuation of the minor axis dust lane prevalent in the
continuum images (Figure~\ref{fig:wiynimages}a). In order to 
estimate the size and orientation of this ring-like
structure, we have fitted it by eye using an ellipse.
The geometry of our fitted ellipse
suggests an approximate 
inclination of $\sim 72^{\circ}$ (assuming the structure 
is intrinsically circular
and has an intrinsic thickness of one-tenth its diameter), and a
position angle of \ad{56}{5} (a misalignment of 8$^{\circ}$ 
from a perfectly polar orientation). 
The diameter of this ellipse ($\sim$2.5~kpc) implies the ring
would lie comfortably inside the bulge of UGC~10043. 
The apparent {\it ansae}
of the ring correspond to the locations where the fitted bulge isophotes
abruptly change position angle and ellipticity, and show
their maximum ``diskiness'' along the minor axis direction
(see Figure~\ref{fig:bulgefits}). 
If real, this ring-like structure may be  related
to the so-called ``inner polar rings'' now identified in 
several nearby galaxies (Reshetnikov, Hagen-Thorn, \& Yakovleva 1995;
Eckart \& Downes 2001;
Karataeva et al. 2001; Sil'chenko 2002; Corsini et al. 2003). 
We estimate the apparent ring structure to have a 
surface brightness roughly 13 times
lower than the adjacent bulge light. To estimate its luminosity, we
summed the flux along the southeastern arc, where bulge and disk
contamination are minimized. Extrapolating this value around
the full ring diameter
yields $m_{R}\approx19.4$, or
$L_{R}\sim1\times10^{7}~L_{\odot}$.

The dusty arc that appears to form part of the 
inner ring structure in UGC~10043
would necessitate that the ring contain dusty, optically thick
material, and suggest its origin might be due to the accretion 
of metal-enriched material from a
companion or infalling dwarf. 
However,
a problem for the inner polar ring scenario is the absence of an
obvious continuation of
the minor axis dust arc to the northeastern portion of the galaxy, as
would be expected from a continuous, orbiting ring. One possible
explanation is that the ring is inhomogeneous, with
the dustiest material confined to discrete clumps. 
Ultimately the inner ring
hypothesis will require kinematic confirmation. As we describe
in Sections~\ref{CaIIkin} \& \ref{ionkin}, the spectra we have obtained
so far do not have sufficient spatial resolution and
signal-to-noise for this purpose.

Lastly, we draw attention to 
the inset shown in Figure~\ref{fig:unsharp}, as it highlights yet another
interesting and unusual feature of the UGC~10043 bulge---namely its
appearance of being
``inserted'' into the disk rather than
forming a smooth, continuous extension of the disk light. 
This contrast is strengthened by a comparison with images or contour plots of other
edge-on spirals with prominent bulges (e.g., de Grijs \& van der Kruit
1996; Howk \& Savage 1999;
Pohlen et al. 2000). 
As discussed below, this distinction may offer another important clue
to the formation history of UGC~10043 (Section~\ref{discussion}).

\subsection{Neighbors to UGC~10043\protect\label{neighbors}}
Several additional galaxies of significant angular size 
are visible within the field-of-view of 
the Mini-Mosaic images of UGC~10043.
One of the most interesting of these
is a very faint, uncatalogued dwarf
at a position $\alpha_{2000}=15^{h}48^{m}38^{s}$,
$\delta_{2000}=+21^{\circ}53^{'}12^{''}$ (roughly $r$=+84$''$,
$z=-30''$), with $m_{B}$=22.1 and $m_{R}$=21.2 
(see Figure~\ref{fig:wiynimagewide}). 
This extremely diffuse and low surface brightness 
object ($\bar\mu_{B}=26.1$ mag arcsec$^{-2}$) has the relatively smooth and
featureless morphology typical of dwarf
spheroidal (dSph) galaxies. 
It does not appear to be a detached fragment from the outer disk of
UGC~10043 itself, as it is located in the anti-direction of the
disk's warp, and has a somewhat redder observed color ($B-R=0.9$)
compared with the
outer disk regions ($B-R\sim$0.7-0.8; see Section~\ref{cmap}). No \HA\
emission was detected from this object in 
our narrow-band imaging observations (Section~\ref{narrowband}), so we
have no quantitative constraint on its redshift. 
However, little or no ionized gas is expected in a dSph system, and
its $B-R$ color and mean
surface brightness are consistent with typical
objects of this class (e.g., Jerjen, Binggeli, \& Freeman 2000); 
assuming this galaxy lies at the same distance as UGC~10043, 
its linear diameter ($\sim$1.1~kpc) 
and absolute magnitude ($M_{B}\approx -10$) are also as expected for
a dSph. As
dSphs are typically satellites to larger
galaxies, this object is a strong candidate 
for a physical companion to UGC~10043. 

Four other galaxies with sizes $\gsim$30$''$ 
were also visible in the WIYN frames.
One of these is MCG~+04-37-035,  at a projected distance of
\am{2}{7}  from
UGC~10043. For it we measure $B$ and $R$ magnitudes of $m_{B}$=16.21 
and $m_{R}$=15.36,
respectively (uncorrected for internal extinction). 
MCG~+04-37-035 is also clearly
detected in our continuum-subtracted 
\HA+[\NII] image (Section~\ref{narrowband}), implying that it has a recessional
velocity in the range $\sim$970-4250~\kms\ (the wavelength range transmitted
by the narrow-band filter), and thus could also be
physically associated with UGC~10043. This galaxy has a rather low
surface brightness with a few small,
brighter star-forming knots superposed. Its disk structure is  
somewhat irregular, but with
hints of rudimentary spiral arms.

A third galaxy in the field was catalogued
by the 2MASS survey\footnote{http://www.ipac.caltech.edu/2mass}, as
2MASX~J15485842+2151508. It lies at a projected distance of \am{4}{0}
from UGC~10043. MCG~+04-37-038, seen \am{5}{1} from
UGC~10043, is catalogued as a galaxy triplet; however, this
designation is puzzling, as the WIYN image clearly shows only a single
galaxy encompassing the positions of all three of the supposed
members of the grouping. Lastly, we find
a previously uncatalogued galaxy at a position
$\alpha_{2000}=15^{h}48^{m}55.4^{s}$,
$\delta_{2000}=+21^{\circ}45^{'}32^{''}$. This latter object, as well as
2MASX~J15485842+2151508, and MCG~+04-37-038, all have similar angular
sizes ($\sim$\am{0}{5}), and no clear evidence of line 
emission detected in the
\HA+[\NII] image. This suggests all three are  background objects and
may be part of a single galaxy group.

\subsubsection{$B-R$ Color Map\protect\label{cmap}}
\begin{figure}
\epsscale{0.95}
\plotone{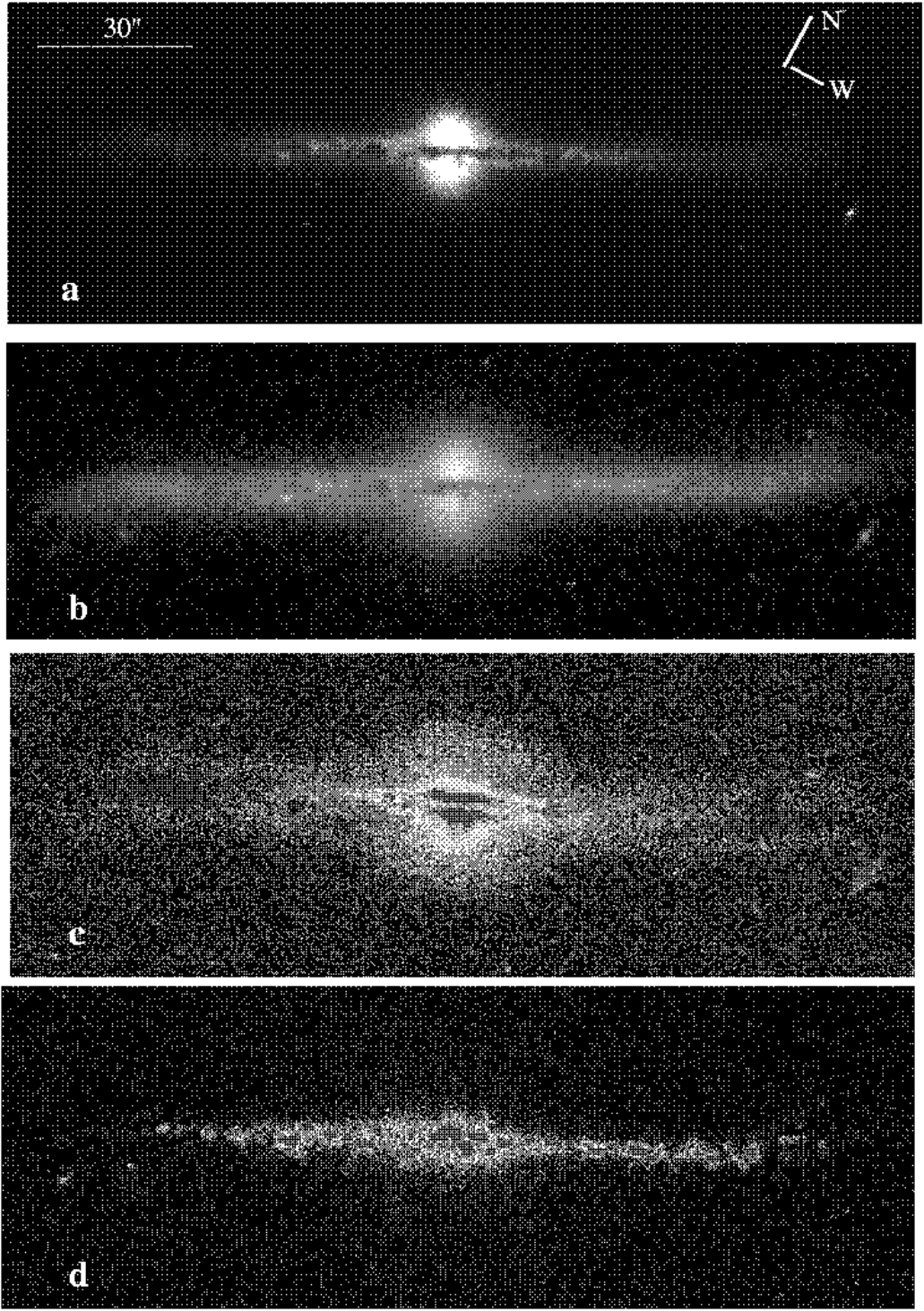}
\figcaption[LDMatthews.fig1_low_bw.ps]{Images of UGC~10043 obtained with the WIYN
telescope: (a) $R$-band; (b) $B$+$R$ composite; (c) $B-R$ color
map; (d)  \HA+[\NII]. 
Seeing was $\sim$\as{0}{6} for the $R$ image and
$\sim$\as{0}{7} for $B$ and \HA+[\NII]. In the color map, dark blue corresponds
to the bluest regions ($B-R\sim$0.5) and red corresponds to regions with
$B-R\sim$2.2; these values are uncorrected for internal reddening.
\protect\label{fig:wiynimages}}
\suppressfloats
\end{figure}

\begin{figure}
\plotone{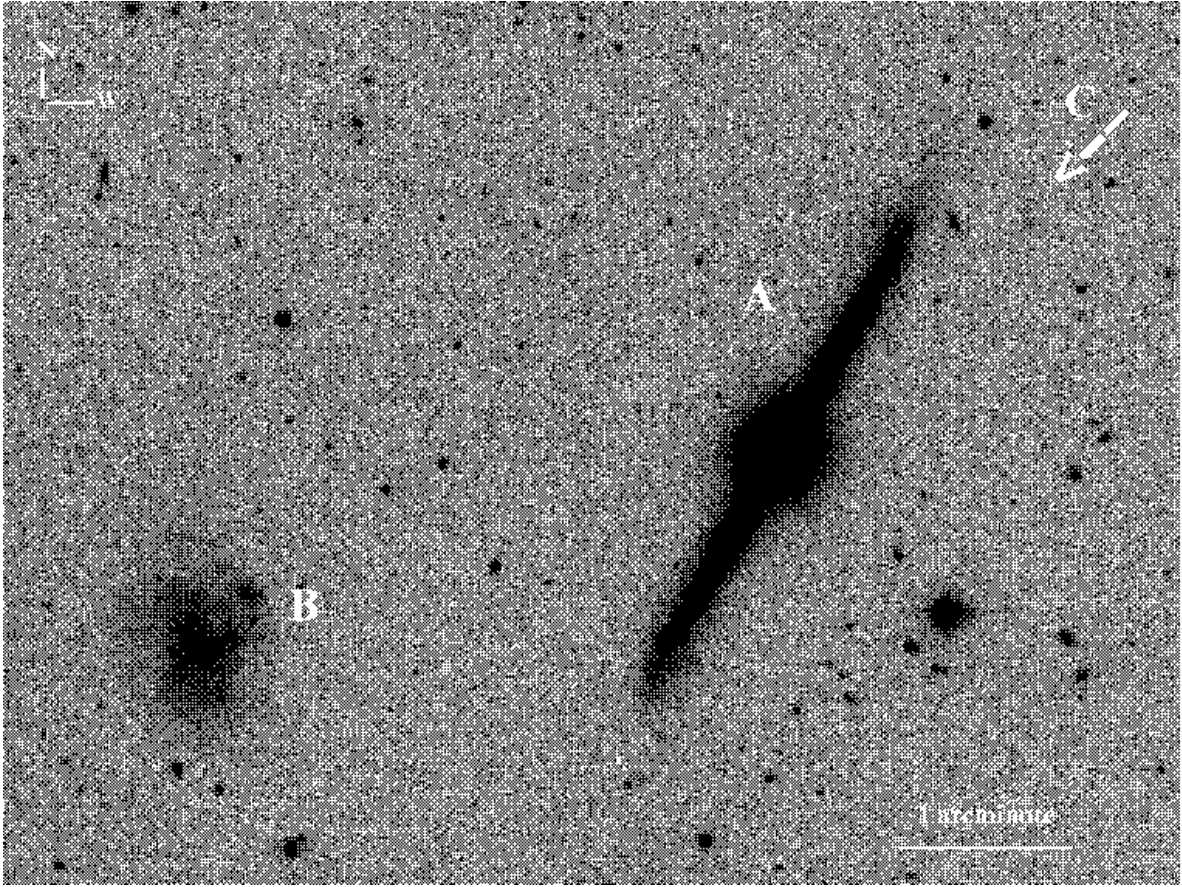}
\figcaption[LDMatthews.fig2_low_bw.ps]{$R$-band image showing UGC~10043 (labelled `A')
and two possible
companions: MCG~+04-37-035 (east of UGC~10043, denoted `B') and a previously
uncatalogued dwarf (to the northwest,
indicated with an arrow and letter `C').  Note the slight
``integral sign'' warp visible in the outer disk of UGC~10043 and the
nearly circular shape of the outer bulge isophotes. This
image is \am{5}{3}$\times$\am{4}{8}. 
\protect\label{fig:wiynimagewide}}
\suppressfloats
\end{figure}

\begin{figure}
\plotone{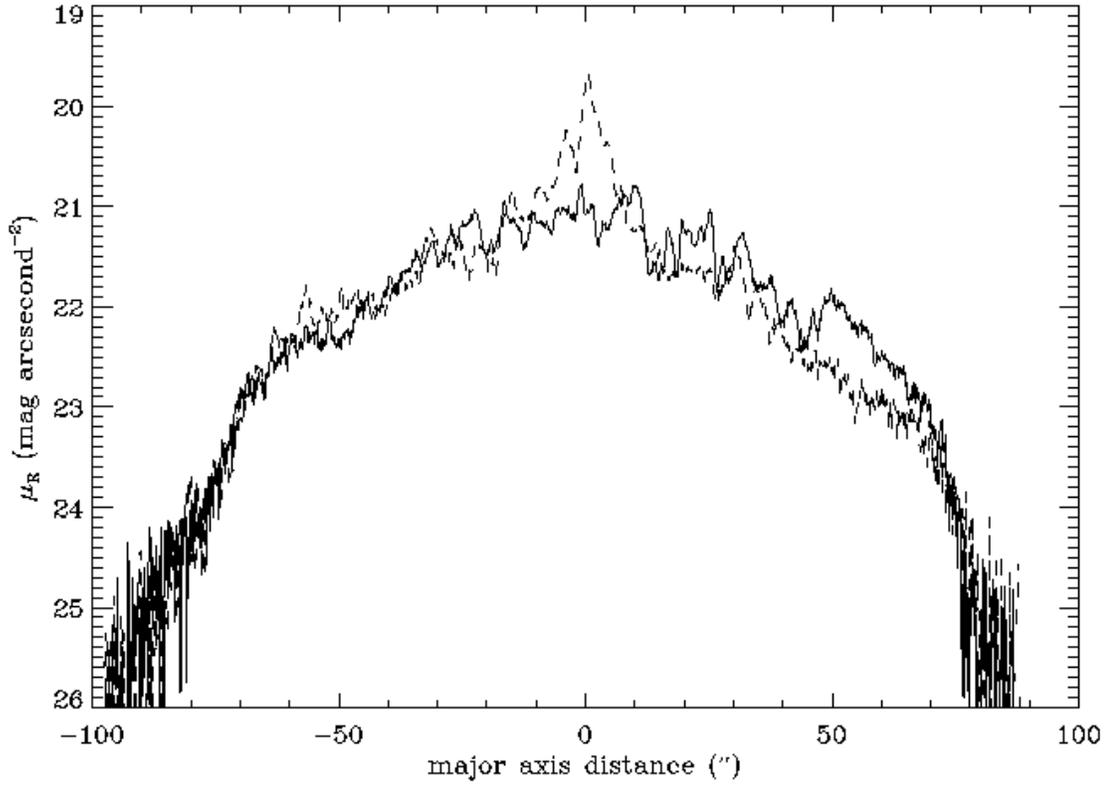}
\figcaption[LDMatthews.fig3.ps]{$R$-band surface brightness 
profiles extracted parallel to the major
axis of UGC~10043. Solid line: major axis profile; dashed line:
profile along $z=+2''$. The profiles were averaged over 4-pixel
($\sim$\as{0}{6}) strips and are uncorrected for projection and
internal extinction. No averaging was applied along the radial
direction. The jagged appearance of the profiles emphasizes the highly
clumped distribution of dust.
\protect\label{fig:Rmajor}}
\suppressfloats
\end{figure}

\begin{figure}
\plotone{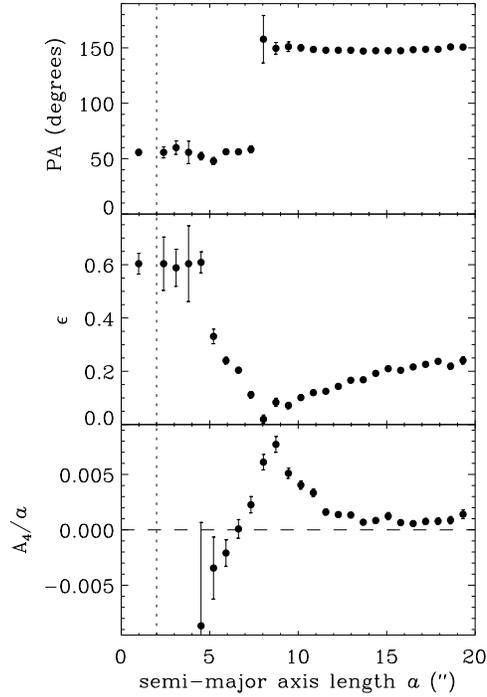}
\figcaption[LDMatthews.fig4.ps]{Results of isophote fitting to 
the bulge of UGC~10043 in
the $R$ band. Various derived quantities are plotted as  function of isophote
semi-major axis:
position angle (top), ellipticity (center) and the ratio
of the Fourier coefficient ${\rm A}_{4}$ to the isophote semi-major axis
(bottom; see text). The vertical dotted line indicates the radius inside which
dust contamination along the major axis is most significant.
Note the abrupt position angle shift near
$a$=\as{7}{5} and the changes in ellipticity as a function of $a$. Because
of the position angle shifts, the elongation of the isophotes inside
$a<8''$ is perpendicular to the disk (prolate), 
while for $a>8''$ the elongation is along the major axis (oblate).
For reference, the disk position angle is
\ad{151}{5}$\pm$\ad{0}{5} (Section~\ref{disk}). 
The change in the ratio of ${\rm A}_{4}/a$ from
negative to positive values implies a transition from boxy to disky isophotes.
\protect\label{fig:bulgefits}}
\suppressfloats
\end{figure}

\begin{figure}
\epsscale{0.90}
\suppressfloats
\plotone{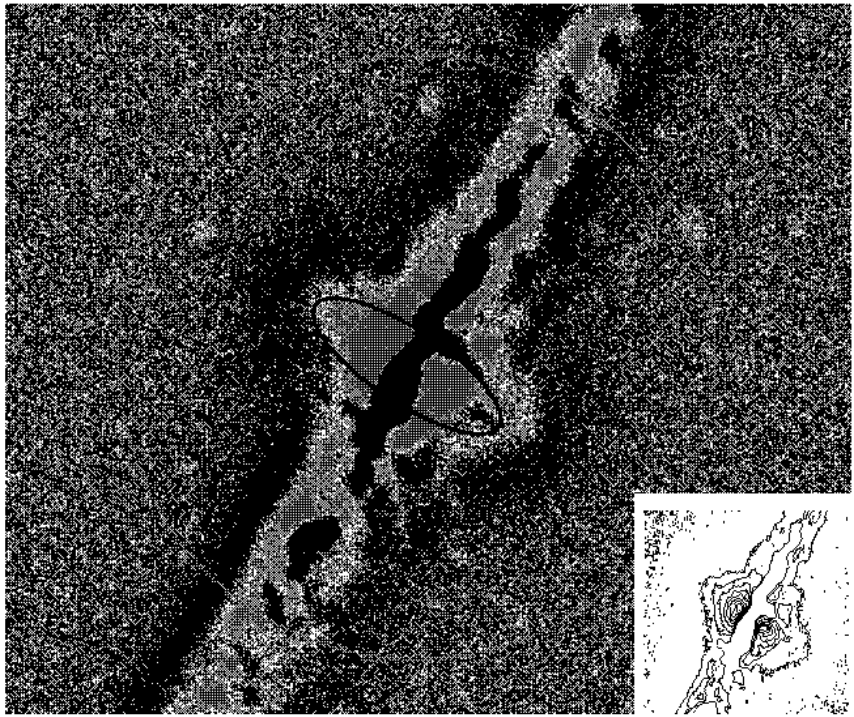}
\figcaption[LDMatthews.fig5_low_bw.ps]{Unsharp mask of the $R$-band 
image of UGC~10043 revealing
a possible inner polar ring structure.
(see Section~\ref{unsharp}). This field is $\sim$\am{0}{9} 
across. The overplotted ellipse (black contour) was used to estimate the size and
inclination of this structure; it has a semi-major axis of 
\as{7}{6}.  The red contour indicates the approximate extent of the
entire bulge. The inset in the lower right corner 
shows an isophotal version of the inner portion of the
same figure. Positive contours from 20-2420 (arbitrary
units) are shown at 300-unit intervals. North is on top, east on 
the left.\protect\label{fig:unsharp}}
\suppressfloats
\end{figure}

A $B-R$ color map of UGC~10043 is presented in
Figure~\ref{fig:wiynimages}c. 
This map reveals several features that provide further 
clues to the evolutionary history of the galaxy.

As seen in the  images shown in
Figure~\ref{fig:wiynimages}a \& b, UGC~10043 forms a stark contrast
to many other dusty, edge-on
disks with prominent bulges (i.e., Hubble types Sb-Sbc; 
e.g., IC~2531: Wainscoat, Freeman, \& Hyland 1989; NGC~891, NGC~3628:
Howk \& Savage 1999). For example,
UGC~10043 does not exhibit a well-defined dust
lane along its full radial extent, but only at small projected radii.
This is reaffirmed 
in the $B-R$ color
map, where we see no color signature of a dust lane outside the
inner $\sim$30$''$  of the galaxy (compare the color maps of edge-on
Sb-Sbc spirals
presented by Wainscoat et al. 1989; de Grijs et al. 1997).
This suggests that the bulk of the dust in UGC~10043 is confined to the inner
several kiloparsecs of the galaxy.
This impression is reinforced 
by the $B+R$ composite image shown in
Figure~\ref{fig:wiynimages}b, although part of this effect may come
from the dust being more readily visible in projection against the bright
bulge light.

\begin{figure}
\suppressfloats
\plotone{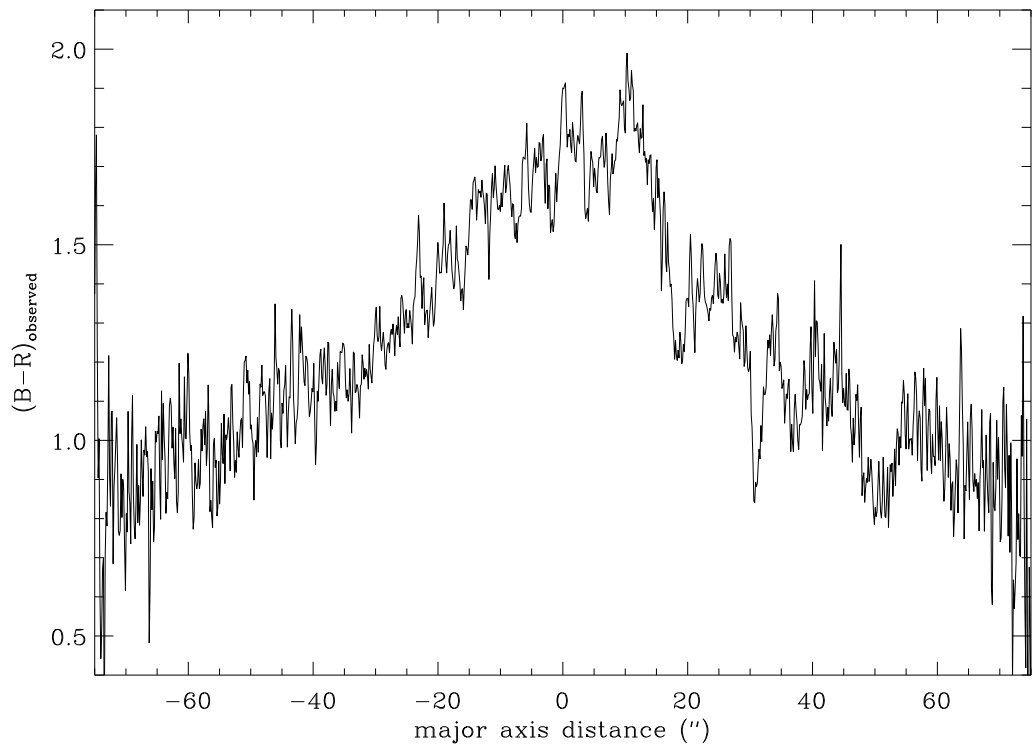}
\figcaption[LDMatthews.fig6.ps]{Observed $B-R$ color profile 
extracted along the major axis of
UGC~10043. 
The data were averaged over a 4-pixel-wide (\as{0}{6})
strip. 
\protect\label{fig:majorcolor}}
\suppressfloats
\end{figure}

A radial color gradient of
$\Delta(B-R)\sim$1.2 mags is observed along the midplane of UGC~10043,
between the center and the edge of the disk, with 
$B-R$ reaching its minimum observed values of $\sim$0.7-0.8 near
$r=75''$ (Figure~\ref{fig:majorcolor}). 
Radial color gradients  (where the disk becomes bluer with 
increasing galactocentric 
radius) are common in the disks of galaxies, and arise from a
combination of stellar population and metallicity gradients (e.g., de
Jong 1996b). If interpreted directly, the large observed
color gradient in the UGC~10043 disk would
suggest that either the  disk built up slowly with
time, and/or that  significant viscous evolution has not 
taken place, perhaps owing to a low surface density
(e.g., Firmani, Hernandez, \& Gallagher 1996; but cf. Bell
2002). Unfortunately, 
radial color  gradients become far more difficult to interpret in edge-on
galaxies (e.g., Matthews \& Wood 2001), 
and three-dimensional radiative transfer modeling  together with
imaging at additional wavebands will be needed to 
disentangle the influences of dust reddening 
and contamination from bulge light
on the color gradient observed in UGC~10043. 

\begin{figure}
\plotone{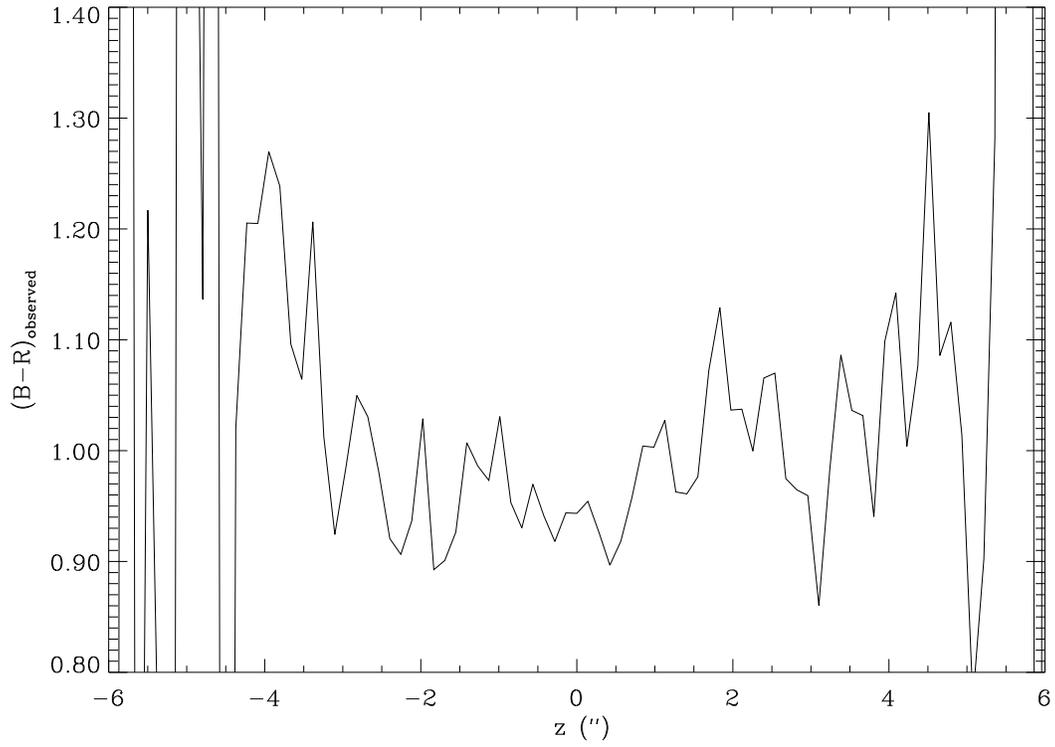}
\figcaption[LDMatthews.fig7.ps]{$B-R$ color profile extracted 
perpendicular to the disk of UGC~10043 at
$r=-60''$. The data were averaged over a 25-pixel-wide (\as{3}{5})
strip. Note the disk becomes redder with increasing $z$-height.
\protect\label{fig:offminorcolor}}
\end{figure}

\begin{figure}
\plotone{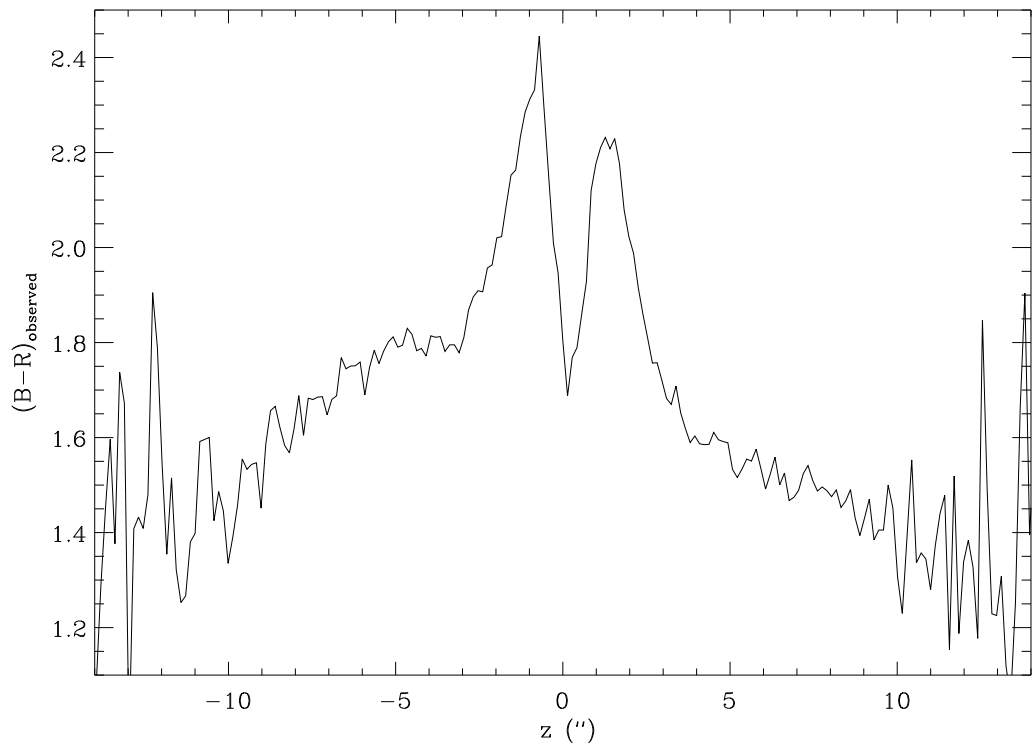}
\figcaption[LDMatthews.fig8.ps]{$B-R$ color profile extracted along the
minor axis of
UGC~10043. The data were averaged over a 3-pixel-wide (\as{0}{4}) strip.
The large dip near $z=0$ is due to the intersecting dust lane.
\protect\label{fig:minorcolor}}
\suppressfloats
\end{figure}

{\it Vertical} color gradients in galaxy disks also provide important clues
to their 
evolutionary histories, as they can unveil the presence of multiple
disk components with different mean stellar compositions and ages (and
hence different characteristic velocity dispersions), as well as
evidence of past dynamical heating (e.g.,
Just, Fuchs, \& Wielen 1996;
Matthews 2000; Dalcanton \& Bernstein 2002, but cf. de Grijs \&
Peletier 2000).  Vertical color gradients
are clearly 
visible in UGC~10043. Outside the bulge region ($r\gsim30''$), 
the disk appears bluer
near $z$=0 than at higher $z$, implying a concentration of the
youngest stars toward the midplane (Figure~\ref{fig:offminorcolor}).
Within the bulge region itself, significant variations in color
are also visible, with a  gradient of 
$\Delta(B-R)\sim$1.0 mags observed along the minor axis over the
interval $|z|=0-14''$
(Figure~\ref{fig:minorcolor}). These bulge color gradients appear to 
arise from a combination of projection effects,
dust reddening, and stellar population differences between the disk and the
inner and outer
bulge. A steep, narrow, blue dip is seen in
the minor axis color profile near $z=0$ (roughly $2''$ wide), 
corresponding to the intersection of the
major axis dust lane. Skirting this are much redder regions with
observed $B-R$
colors as red as 2.4;
these latter values are $\sim$0.8 magnitudes redder 
than the reddest dust-corrected
$B-R$ colors of old, composite 
stellar populations (Caldwell \& Rose 1997), and
imply significant reddening from dust must be occurring to $z$-heights
at least $\sim 3''$ from the midplane. 

To explain the
complex distribution of colors in the central regions of UGC~10043, we propose
the following picture:  near $z$=0, 
dust within the disk located at intrinsically small $r$
is sufficiently optically thick that from our viewing angle, 
the bulge light is entirely 
occulted at $B$ and $R$ wavelengths, and 
we see only the bluer, outer disk
light projected in front of it. As the optical depth of the inner disk drops with
increasing $z$, the  bulge light at slightly higher $z$
is able to reach us, but only after being
substantially reddened. 
Given the relatively small scale height of the disk, 
beyond $|z|\gsim3''$, the bulge light is likely to be minimally 
contaminated. On the
southwest side of the midplane, 
the bulge light near the minor
axis is further
reddened by the perpendicular dust structure discussed above (see
Section~\ref{optmorph}, 
Figure~\ref{fig:wiynimages}a \& b). Finally, beyond 
$|z|\gsim10''$ the color gradient on both sides of the bulge
appears to flatten off to a fairly
uniform value of $B-R\sim$1.4.  
The modest color gradient that is observed
over the interval $|z|\approx 4-10''$ [$\Delta(B-R)\sim$0.3] 
is similar to radial color gradients observed in
the bulges of other Sbc galaxies (Balcells \& Peletier 1994), and may be
due to a modest stellar population/metallicity gradient. 
Interestingly, no abrupt color shifts are seen corresponding with the isophote
position angle shifts discussed in Section~\ref{bulge}.

One additional noteworthy feature of Figure~\ref{fig:wiynimages}c is
the color structure of the inner disk regions, where we see evidence
that the inner disk ($r\lsim30''$)  is tilted
relative to the outlying disk by roughly 3-5$^{\circ}$
(note the regions appearing 
as green and yellow in Figure~\ref{fig:wiynimages}c). 
Further evidence that this component is intrinsically tilted 
comes from the observation that 
the apparent midpoint of the major axis dust lane (presumed to arise
from dusty material at small or intermediate $r$) does not
lie on a straight line with respect to an extrapolation of the
midplane from the outer
disk regions (see Figures~\ref{fig:wiynimages}a \& b).  
Typically, the presence of a tilted inner disk structure is
difficult to infer in an edge-on galaxy due to projection and optical
depth effects. However, other examples of this phenomenon are seen
in the Milky Way (e.g., Liszt \& Burton 1980; Sanders,
Solomon, \& Scoville 1984), the polar disk galaxy NGC~4650A
(Gallagher et al. 2002), and the edge-on S0/Sb spiral NGC~5084 (Zeilinger,
Galletta, \& Madsen 1990). One possible origin for these types of
tilts has been proposed by Ostriker \&
Binney (1989); they suggest that a continual slewing of the galactic
potential due to infalling material could produce both inner-disk
tilts, as well as outer-disk warps (see also
Debattista \& Sellwood 1999). 

\subsection{Narrow-Band Imaging\protect\label{narrowband}}
A continuum-subtracted 
\HA+[\NII] image of UGC~10043 is presented in Figure~\ref{fig:wiynimages}d.
A smooth  component of ionized gas
is visible throughout the UGC~10043 disk, together with a number of
brighter \HII\ complexes. In the outer disk regions, the scale height
of the \HA\ emission appears smaller than that of the $R$-band
continuum emission, and the diffuse component is rather weak. 
On the northwestern 
tip of the disk, line emission is seen extended down to $z=$\as{12}{5}
($\sim$2~kpc)  from
the midplane, tracing the stellar warp
(see Section~\ref{disk} and
Figure~\ref{fig:wiynimagewide}). The southeastern 
side of the disk exhibits somewhat less 
emission at large galactocentric radii than the northwestern side,
but two moderately bright \HII\ regions
are found displaced roughly 7$''$ (1.1~kpc) 
from the plane, together with a few
faint patches of diffuse emission; these emission regions trace the outer
envelope of the southeastern edge of the 
stellar warp. 

The most striking feature of Figure~\ref{fig:wiynimages}d
is the structure of the emission in the
central regions of UGC~10043. The ionized gas layer appears noticeably
thicker in the inner $\sim$30$''$ 
($\sim$5~kpc) of the galaxy, as traced by both the
$z$-extent of \HII\ complexes and the more smoothly distributed
component. Moreover, faint, diffuse emission can be traced
as far as $|z|\approx20''$ ($\sim$3.2~kpc) 
above and below the plane (at $\ge$2.5$\sigma$), forming a roughly 
biconical or ``hourglass''-shaped structure.  To further highlight the
morphology of this low surface brightness, 
vertically-extended emission, we have convolved the image shown in
Figure~\ref{fig:wiynimages}d with a Gaussian with $\sigma$=2.5 pixels,
yielding in a FWHM resolution of $\sim6''$.
The resulting image is shown in Figure~\ref{fig:HAsmooth}.

``Hour-glass''-shaped
distributions of ionized gas with size scales of a few kiloparsecs or more
are a hallmark of galaxies possessing large-scale galactic winds
(e.g., Heckman, Armus, \& Miley 1990;
Lehnert \& Heckman 1996; Veilleux \& Rupke 2002). 
Figure~19 of Heckman et al. (1990) shows a cartoon sketch 
of the geometry of
such a wind. These types of biconical winds are believed to be 
powered by a combination of multiple supernovae
and stellar winds within regions of intense, centrally-concentrated star formation.
The supernovae and stellar winds are expected to
create a hot cavity that expands into an ambient halo
medium, dragging along disk material (e.g., Suchkov et al. 1994). 
As we discuss in Section~\ref{radio}, there are
several independent pieces of evidence for enhanced nuclear star
formation  in UGC~10043, albeit somewhat less than typifies most
galaxies known to have large-scale winds.
Moreover, the spectroscopic observations that we describe in
Section~\ref{ionkin} provide additional confirmation of
the presence of expanding or outflowing 
gas and vertically increasing [\NII]/\HA\ ratios
in UGC~10043, as expected in the presence of a
galactic wind. 
We further discuss the properties of the wind 
and the implications for the evolutionary history of UGC~10043
in Sections~\ref{ionkin} \& \ref{discussion}.

\section{Spectroscopic Observations and Data 
Reduction\protect\label{spectroredcut}}
Gaining further insight into the complex structure of the UGC~10043
system requires kinematic information to complement the imaging
studies described above.
In order to obtain two-dimensional kinematic measurements of the
galaxy, we employed the DensePak integral field
spectrograph on the WIYN telescope. 
DensePak is a fiber spectrograph that consists of ninety-one, 3$''$-diameter
optical fibers arranged in a 7$\times$13 array. Five of the array
fibers have failed since assembly.   The array samples a 
45$''\times27''$ region on the sky, with \as{4}{15}
separation between fibers. Four additional fibers
can be used to obtain simultaneous sky spectra. A more detailed
overview of DensePak can be found in Barden, Sawyer, \& Honeycutt (1998).

With DensePak we obtained observations of UGC~10043 over the
wavelength range containing the redshifted \HA\
$\lambda\lambda$6562.8\ang\ and [\NII] 
$\lambda\lambda$6548.0,6583.4\ang\ lines in
order to
probe the kinematics of the ionized gas, and over the wavelength 
range containing the
\CaII\ infrared triplet ($\lambda\lambda$8498.0,8542.1,8662.1\ang) in order
to measure the stellar kinematics. 
These spectra were obtained on 2001 April 14 and 15,
respectively. Seeing was $\sim$\as{0}{7}-\as{0}{8} 
during the two nights. On each night,
DensePak was positioned on the galaxy visually 
using the TV acquisition camera, hence the absolute position of the
array center is known to an accuracy of roughly a few
arcseconds. A position angle of $0^{\circ}$  was used for the fiber array
(i.e., no rotation), resulting in the long axis of DensePak aligned
north-south, and
the galaxy major axis lying roughly along the diagonal of
the array. The locations of the array fibers during the observations are
indicated on $R$-band and \HA+[\NII] images of UGC~10043 in
Figure~\ref{fig:denseoverimages}. 
As seen from this figure, the array sampled the bulge and
the inner disk regions of the galaxy.

The DensePak fibers were fed to the Bench Spectrograph
Red Camera, which employs a 2048$\times$2048 T2kC thinned CCD with 24$\mu$m
pixels, a gain of 1.7~$e^{-}$ ADU$^{-1}$, and a read noise of
4.3~$e^{-}$. For the observations covering the \HA\ region, 
the 860~$l/{\rm mm}$ grating was used in second order, 
with the G5$\_$GG-495 
blocking filter. The center wavelength was 6280\ang, 
resulting in a wavelength coverage of 5811-6757\ang\ and a dispersion
of 0.461\ang\ 
($\sim$22 \kms) per pixel. Gaussian fits to unblended 
night sky lines
yielded an estimate of the 
FWHM resolution of $\sim$59~\kms. For the \CaII\ spectra, the
860~$l/{\rm mm}$ grating in first order was used, centered at 8700\ang, 
together with the G2$\_$RG-695 filter. This 
yielded a wavelength coverage of  
7543-9701\ang\ and a dispersion of 1.053\ang\ ($\sim$36 \kms) per
pixel. FWHM resolution, as determined by fitting a Gaussian to several
night sky lines, was $\sim$86~\kms.

Several bias and dome flat exposures were taken at the beginning of each
night and combined  using IRAF. During the \HA\
observations, spectra of CuAr lamps were taken regularly throughout
the night to provide wavelength calibration. 
For the \CaII\
observations, the copious night sky lines were used for 
calibration, with wavelengths for the sky lines 
taken from Osterbrock, Fulbright, \&
Bida (1997).
Total integration times on the galaxy were 3$\times$1200 seconds in
the \HA\ + [\NII]
region and 4$\times$1200 seconds in the \CaII\ region.

Reduction of the spectra began with the subtraction of overscan and bias
levels from all frames. Next, the 
individual galaxy exposures and comparison lamp exposures 
were combined and cleaned of cosmic rays using the `crclean'
algorithm provided by M. Bershady. The remainder of the reduction, 
including  
wavelength calibration, dispersion correction, sky subtraction, and spectral
extraction, was performed using the 
`hydra' package within IRAF. 
Median dome flat exposures (obtained using a quartz lamp) were used to
define the extraction apertures and to correct the spectra for
fiber-to-fiber sensitivity variations.
 
For the \HA\
data, sky subtraction was achieved using the four designated sky fibers
(see above). For the \CaII\ spectra, since the galaxy lines were not
detected in a number  of the ``object'' fibers,
(see Section~\ref{triplet} below), 10 additional line-free
fibers were averaged to improve the signal-to-noise of the sky subtraction.
RMS dispersions in the final wavelength solutions were 0.026\ang\ 
($\sim$1.2~\kms)
for the \HA\ spectra and 0.043\ang\ ($\sim$1.5 \kms) for the \CaII\ spectra.

\section{Analysis of the Spectroscopic Data\protect\label{spectroscopy}}
\subsection{Stellar Kinematics from the \CaII\ Triplet\protect\label{triplet}}
Absorption lines from the \CaII\ infrared triplet 
were unambiguously detected in UGC~10043 in 31 of the 86
working DensePak fibers. 
The detection of these lines was limited to the bulge and a few points
along the disk. Those disk locations where \CaII\ was detected
correspond with the higher surface brightness
``inner disk'' region prevalent in the
$B-R$ color map of Figure~\ref{fig:wiynimages}c.
Some sample \CaII\ spectra are shown in
Figure~\ref{fig:CaIIsamples} to illustrate the typical quality of the data. 
The \CaII\ spectral region is contaminated by numerous telluric OH
emission lines as well as the broad atmospheric A-band (O$_{2}$) 
emission feature near 7594\ang. 
Residuals of these lines often remained  after sky subtraction,
making it difficult to determine the continuum
levels in this wavelength region with high accuracy. 
This had little effect on the line
centroid measurements, but did add uncertainty to the inferred stellar
velocity dispersions (see below).
In several of the  spectra, the \CaII\
lines show hints of being comprised of multiple components (see 
Figure~\ref{fig:CaIIsamples} for an example). This cannot be
attributed to night sky emission residuals, as fortuitously, no
night sky emission lines overlap with the redshifted \CaII\ lines
from the galaxy. Where  \HII\ regions are present, nebular Paschen
series lines can be present in this region of the spectrum, including
three that overlap in wavelength with the \CaII\ triplet (see Terlevich et
al. 1996); however we see no evidence of lines from this series
elsewhere in any of our spectra, suggesting Paschen emission is not causing
significant contamination of the \CaII\ absorption profiles.
Unfortunately, the signal-to-noise of the present data 
is insufficient to permit well-constrained, 
multiple-component
fits to the individual lines, and 
higher resolution, higher signal-to-noise data will be needed to
confirm if these multiple components are real and to better constrain
their physical origin.

In order to derive the stellar kinematics from the \CaII\ lines, we
cross-correlated the spectra from each fiber against a synthetic
template of a zero rotational velocity 
K1.5 giant from Kurucz (2003), broadened  to match the 
instrumental resolution of the observations.  The model template was
oversampled and therefore effectively 
noise-free. In addition to line-of-sight radial velocity measurements, 
the cross-correlation yielded velocity 
dispersion measurements from the widths of the correlation peaks (see
Section~\ref{veldisp}).
The errors on each quantity were derived following the
prescription of Tonry \& Davis (1979), and do not take into account
errors arising from possible non-Gaussian shapes of the lines.

The derivation of true rotational velocities in the galaxy would
require corrections to the observed 
velocities for line-of-sight integration effects and
asymmetric drift. However, both corrections are expected to be small 
compared with the size of the formal and systematic 
errors of the present data, and the uncertainties in the assumptions
required for applying these
corrections are likely to be comparable to the size  of these
corrections themselves (see e.g., Whitmore, McElroy, \& Schweizer
1987). 
Therefore for the discussion that follows, we
simply utilize the observed, line-of-sight radial velocities. That the
asymmetric drift corrections are not large is underscored by the good
agreement between the \CaII\ velocities and the velocities inferred
from the \HA\ and [\NII] emission lines at a given
$r$ (see Figure~\ref{fig:NIImajor}, discussed below).

\subsubsection{$\CaII$ Line-of-Sight Radial Velocities and the
Kinematics of the Bulge\protect\label{CaIIkin}}
A map of \CaII\ line-of-sight radial velocities as a 
function of fiber position
is shown in Figure~\ref{fig:bmontage}a. Error bars range from
7-20~\kms\ in most cases (23 fibers), but are as high as 20-40~\kms\ in
a few fibers. 

As seen in Figure~\ref{fig:bmontage}a, some evidence of a velocity gradient is
apparent along the galaxy major axis, 
indicative of rotation (see also Figure~\ref{fig:NIImajor}, discussed
below). The peak \CaII\ rotational velocities observed along
the fibers closest to 
major axis are $\sim\pm$70~\kms, and occur at the last measured
points near $r\approx\pm15''$.  However, the bulge model derived in
Section~\ref{optmorph} and the apparent high optical depth
in the midplane at small $r$ (e.g., 
Figure~\ref{fig:minorcolor}), both suggest that the two rows of 
fibers nearest the midplane 
are likely to be significantly contaminated by disk light (see also
Figure~\ref{fig:denseoverimages}).
This implies that the observed rotation is not necessarily
intrinsic to the bulge itself. Indeed at higher $z$-heights
($|z|\gsim4''$), no 
evidence of systematic 
rotation is seen, suggesting the bulge is rotating  slowly,
if at all, about the galactic pole.

Lack of significant bulge rotation is unusual for 
a normal Sbc galaxy, where frequently bulge rotation
speeds are
comparable with those in the adjacent disk (e.g., Kormendy \& Illingworth 1982;
Kormendy 1993). However, in the case of UGC~10043,
it seems congruous both with the lack of
significant flattening of the bulge isophotes and with the apparent
geometric decoupling between disk and bulge (e.g., Figure~\ref{fig:unsharp}).
In addition, this behavior is consistent with other
early-type spirals with orthogonally elongated bulges (Bertola et
al. 1999; Sarzi et al. 2000; Reshetnikov et al. 2002; Corsini et
al. 2003), as well as with polar ring galaxies, where the 
``bulge'' is of course actually a subsystem rotating {\it 
orthogonally} to the disk plane. 
Therefore, given the peculiar isophotal structure of the UGC~10043 bulge,
a key goal of our spectroscopic measurements 
was to investigate the bulge kinematics of UGC~10043 
{\sl perpendicular} to the disk in order to search for similar signs
of orthogonal rotation.

In Figure~\ref{fig:CaIIminor}, we plot the derived 
radial velocities (relative to $V_{sys}$)
as a function of $z$-distance  for all
fibers within $|r|<5''$ where the \CaII\ lines were detected
(fibers 29,30,35,37,38,41-45,48-50,56). 
Because of the major axis gradient discussed above, 
the values near $z=0$ show an intrinsic
spread, depending on whether they are measured on the approaching or receding
side of $r=0$. To compensate for the observed scatter and to
best emphasize any underlying systematic
gradient along the $z$ direction, we computed at each $z$ value an
(unweighted) mean velocity
for each group of measurements. Based on these mean values, the data in
Figure~\ref{fig:CaIIminor} reveal 
no evidence for a rotational gradient along the minor
axis of UGC~10043.  

Unfortunately, the rather coarse spatial 
sampling provided by DensePak, coupled with the existence of two
failed fibers near the center of the array, mean that relatively few
fibers sample the bulge of UGC~10043, particularly the inner,
vertically elongated region, which is only $\sim$15$''$ across. As a
result, kinematic peculiarities confined to the inner few arcseconds
of the bulge ($r\lsim5''$) would likely be undetected. 
Another complication in interpreting the \CaII\ data is that, as noted above, a
number of the spectra show hints of multiple velocity
components that cannot be reliably decomposed due to insufficient 
signal-to-noise. Nonetheless, the present data seem to rule out 
any significant global minor axis rotation of either the entire
UGC~10043 bulge, or its entire
vertically-elongated inner component. This result implies that the peculiar bulge
of UGC~10043 is inconsistent with an S0-like galaxy tilted orthogonally to the main
disk, and that UGC~10043
is not a classic polar ring galaxy masquerading as an
edge-on spiral.

That UGC~10043 is different in several important respects 
from other kinematically-confirmed polar
ring galaxies was already indicated by the imaging data discussed above.
An S0-like galaxy would not be expected to show the type of isophote twisting
and oblate outer isophotes seen in the bulge of UGC~10043
(Section~\ref{bulgedecom}). In addition, in typical polar ring systems, the
central (S0-like) component typically 
dominates the total light of the galaxy, unlike
the case for UGC~10043 (Section~\ref{bulgedecom}). 
Finally, photometrically there is no
evidence that the stellar
disk of UGC~10043 is actually a ring or annulus, since we see no sign
of edge-brightnening (e.g.,
Figure~\ref{fig:Rmajor}). The presence
of a large-scale galactic wind (Section~\ref{ionkin}), seemingly
emulating from the inner disk regions also argues against this
picture. Nonetheless, we wish to emphasize that a disk rather than a
ring morphology in itself does not exclude the possibility of an
evolutionary link with certain types of polar ring galaxies. 
There is now evidence that some polar ``ring'' systems, 
including a prototype of this class,
NGC~4650A, actually have a complete or nearly-complete {\it polar disk} 
(e.g., Gallagher et al. 2002;  Iodice et al. 2002). Another strong
candidate for
a ``polar disk'' galaxy is 
ESO~603-021, whose peculiar ``bulge'' shares a number of
intriguing structural similarities with UGC~10043 (Arnaboldi, Capaccioli,
\& Combes 1994; Reshetnikov et
al. 2002). Recent numerical
work suggests systems like NGC~4650A and ESO~603-021
may have a different formation mechanism from polar ring galaxies
with narrow rings (e.g.,  Bekki 1998; Bournaud \& Combes 2003). Moreover,
certain similarities between
UGC~10043 and `polar disk' galaxies suggest it 
may provide some important new clues for understanding the origins of
these latter systems
(see further discussion in Section~\ref{discussion}).

As discussed in the Introduction, 
in addition to the kinematically-confirmed ``polar disk'' galaxies,
a handful of other examples of bona fide disk galaxies are now known that
have vertically
elongated and/or orthogonally decoupled bulges. However, these systems are
distinct from polar disk galaxies in that their entire ``bulge''
does not exhibit minor axis rotation, but instead it contains a small,
orthogonally-rotating inner core, only a few
arcseconds across
(Bertola et al. 1999; Sarzi et al. 2000; Corsini et
al. 2002,2003). 
One of the first discovered examples of this class is 
the Sa/Sab galaxy NGC~4698 (Bertola et
al. 1999). NGC~4698 is seen at a somewhat lower inclination than UGC~10043
($i=65^{\circ}$), but it too exhibits a bright bulge that appears visually
elongated in a direction perpendicular to its lower surface brightness, dusty disk
(see Sandage \& Bedke 1994).
A parametric decomposition by Bertola et al. (1999) shows the
bulge to be elongated over its full extent ($r\sim 50''$), although
orthogonal rotation (with peak amplitude $\sim$30~\kms)
is detected only within a small, inner core
($|r|<6''$). This ``core'' was later found by Pizzella et
al. (2002) to be a nuclear disk. 
In the preferred interpretation of Bertola et al. (1999), 
no rotation of the bulge occurs along the disk major
axis of NGC~4698.
However, as discussed by those authors,
an independent, non-parametric decomposition of the galaxy light by
Moriondo, Giovanardi, \& Hunt (1998)  implies much rounder bulge
isophotes ($\epsilon$=0.08). With this model, a self-consistent picture of the
photometry and kinematics
would then demand some rotation of the bulge along the major axis, plus
an orthogonally rotating core, 
together with a third central luminous component (see
also the discussion in Sarzi et al. 2000). 
This latter picture appears to be 
closer to the one emerging for UGC~10043.

Unfortunately, because of the limited spatial resolution of the
DensePak data, we cannot yet confirm whether or not UGC~10043 may also
have a
small ($r\lsim 5''$), kinematically decoupled inner core or nuclear disk. 
The search for the kinematic signatures of 
such a component using high-resolution longslit
spectroscopy will be an important next step both in 
establishing whether such cores are ubiquitous in galaxies with
vertically elongated, non-rotating bulges,
and also in uncovering the
formation history of these types of galaxies. 

Although our \CaII\ spectra do not reveal
any clear evidence of orthogonal bulge rotation, 
they do reveal one potentially important clue to the overall bulge
structure of UGC~10043 in the form of  apparent 
velocity shifts of $\sim -$30~\kms\ on either side of the bulge. 
Although the statistical
significance of this effect is only marginal, these velocity shifts
have the same sign on both sides of the bulge and are seen
to correspond precisely with the
photometrically-defined transition between the inner and
outer bulge isophotes (Section~\ref{bulgedecom})---i.e., with the
location where
the bulge isophotes abruptly change ellipticity
and position angle (denoted by arrows in 
Figure~\ref{fig:CaIIminor}).
Kinematic signatures of this types are expected in
triaxial potentials at the locations where orbits transition from elliptical to
circular (e.g., 
Corsini et al. 2003). This therefore adds possible {\it kinematic} evidence
to the case for a triaxial bulge in UGC~10043 (see also Section~\ref{bulgedecom}).

\subsubsection{Stellar Velocity Dispersions\protect\label{veldisp}}
Stellar velocity dispersions ($\sigma_{*}$) inferred from the cross-correlation
analysis of the \CaII\ lines
are plotted as a function of fiber number in
Figure~\ref{fig:veldisp}. Typical $\sigma_{*}$ 
values are found to lie in the range 100-130~\kms. Overall, 
the formal uncertainties on the velocity
dispersion measurements are rather large, due to a combination of
residual contamination from night sky lines (which adds uncertainty to
the continuum level determination), and the
fact that line broadening due to random stellar motions in UGC~10043
is small
compared to the instrumentally-broadened 
\CaII\ line widths of the K-giant star
used as a template for the cross-correlation.
We find no evidence of systematic changes in 
the $\sigma_{*}$ values as a function of position in UGC~10043,
although given the large error bars, the presence of modest
vertical or radial gradients in $\sigma_{*}$  cannot be ruled out. 

\subsubsection{$V_{{\rm max}}/\sigma_{*}$-$\epsilon$ and
$L\propto\sigma^{n}_{*}$ Correlations\protect\label{faberjackson}}
Given the complex structure of the UGC~10043 bulge, it is of
interest to see how some of its characteristic parameters compare with
established relations for spiral bulges and other spheroidal systems.
Taking as a representative value of the stellar velocity dispersion the mean
value for the 10 points with the smallest error bars
($<$40~\kms) gives $\bar\sigma_{*}=113\pm13$~\kms.  As discussed in
Section~\ref{CaIIkin}, if we consider
only the fibers with $|z|>4''$ (where disk contamination is
minimized), 
our current data show no compelling evidence of
significant rotation of the bulge along the galaxy major axis,
suggesting $V_{{\rm rot}}<20$~\kms. This implies the UGC~10043 bulge is
characterized by a ratio $V_{{\rm max}}/\sigma_{*}<0.2$. 
Adopting a
representative $\epsilon$ value for the outer bulge isophotes of
$\epsilon\sim$0.2 (see Figure~\ref{fig:bulgefits}), we can then
place the bulge of UGC~10043 on the 
$V_{{\rm max}}/\sigma_{*}$ versus $\epsilon$ diagram for spheroidal
systems. This relation is of interest for spheroidal systems,
since under the simple 
assumption that spheroids are oblate with isotropic velocity dispersions,
the virial theorem predicts a fixed relationship between these two
quantities. Indeed, the bulk of early-type bulges are seen to adhere
fairly closely to this
relation (e.g., Kormendy 1993, his Figure~3). However, in the case of
UGC~10043, the observed outer-bulge value of $\epsilon$ would predict a ratio
$V_{{\rm max}}/\sigma_{*}\sim0.5$. In contrast, the observed value of 
$V_{{\rm max}}/\sigma_{*}<0.2$ places UGC~10043 in the portion of the 
$V_{{\rm max}}/\sigma_{*}$-$\epsilon_{*}$ plane 
inhabited by elliptical galaxies rather than bulges (see also Kormendy \& 
Illingworth 1982; Davies et al. 1983).
 
We can also compare the $\sigma_{*}$ and $M_{R}$
values derived for the UGC~10043 bulge
with standard $L\propto\sigma^{n}_{*}$ (Faber-Jackson) relations. 
Adopting the $R$-band bulge luminosity
determined in Section~\ref{optmorph}, assuming a mean color for the
bulge region of $\overline{(B-R)}=1.5$, 
and assuming internal extinction in the bulge to be negligible, the $B$-band
$L\propto\sigma^{n}_{*}$ relation for SA0-Sbc bulges 
derived by Kormendy \& Illingworth (1983) predicts a central
velocity dispersion of $\sigma_{*}(0)\sim$140~\kms. Their analogous
relation derived from a sample of elliptical galaxies predicts 
$\sigma_{*}(0)\sim$124~\kms. The value measured
in the center of UGC~10043 (fiber 43) is
$\sigma_{*}=117\pm35$~\kms---consistent with either prediction to
within observational uncertainties. 

\subsection{Kinematics of the Ionized Gas\protect\label{ionkin}}
As described above, a second set of DensePak observations of UGC~10043
focussed on the wavelength range encompassing the 
\HA\ and [\NII] emission lines in order to permit an investigation of the 
kinematics of
the ionized gas. Figure~\ref{fig:denseoverimages} shows
the adopted positioning
of the DensePak array on an \HA+[\NII] image of the galaxy. In total,
\HA\ emission was detected (at $>2\sigma$) in 68 of the 86 functioning DensePak
fibers and [\NII] was detected in 63 fibers.

Small portions of the \HA+[\NII] spectra  are plotted as a function
of position in
Figure~\ref{fig:HAgrid} in order to give a sense of how the line
shapes, widths, and ratios change as a function of location in the
galaxy. Larger versions of
a few representative spectra are shown in
Figure~\ref{fig:HAsamples}. It is evident from these figures that in many
cases, both the
\HA\ and [\NII] lines
resolve into multiple velocity components. Complex
line profiles are especially prevalent above and below
the midplane (e.g., fibers 13, 29, \& 57), although a number of the
lines observed along the midplane also show hints of complex
structures  (e.g., fibers 43 \& 65). We further discuss the
physical implications of the line shapes below.

In order to measure the velocity field traced by the \HA\ and [\NII]
emission, we fitted
Gaussians plus a linear background
to the emission lines in each fiber. Owing to the edge-on geometry of
UGC~10043, 
lines-of-sight toward the inner regions of the galaxy intersect material at
a wide range of intrinsic galactocentric radii. Consequently, the
observed line shapes  are not
expected to be purely Gaussian, but rather to have a
Gaussian envelope on the side of the velocity extremum, plus
a tail toward the systemic velocity (see Figures~2 \& 3 of 
Garc\'\i a-Ruiz, 
Sancisi, \& Kuijken 2002 for an heuristic illustration). 
The result is that at small or intermediate $r$,
intensity-weighted Gaussian fits will
tend to underestimate the true rotational velocities and may produce a
rotation curve somewhat shallower than the intrinsic disk rotation
curve. However, because of the
complex structures of the \HA\ and [\NII] line
profiles in UGC~10043 and the limited spectral resolution
of the current observations, 
we have not attempted to correct for this effect.  
This does not affect any of the conclusions that follow.

Initially, we fit only a single Gaussian to each emission
line in order to obtain a global impression of the kinematics. 
No constraints on the widths or
relative velocity offsets of the fitted components were imposed. We
fitted both lines of the [\NII] doublet to supply a check on wavelength
calibration and error estimates, but
for the analysis
that follows, we use the [\NII] velocities derived for the
stronger line in the doublet. Uncertainties on the line centroids were
estimated as $\sigma_{\rm err}=RS^{-0.5}$~\kms, where $R$ is the velocity
resolution per pixel and $S$ is the signal-to-noise ratio of the 
line. $S$ was evaluated by taking the
ratio of the peak flux in the fitted line to the spectral
rms (evaluated over the line-free portion of the spectrum
between roughly 6500-6700\ang). Estimated errors for the \HA\ 
velocities range from 3-17~\kms, with a mean value $\sim$9~\kms; for
the [\NII] velocities, errors range from 5-27~\kms, with a mean of $\sim$10~\kms.

\begin{figure}
\epsscale{0.7}
\plotone{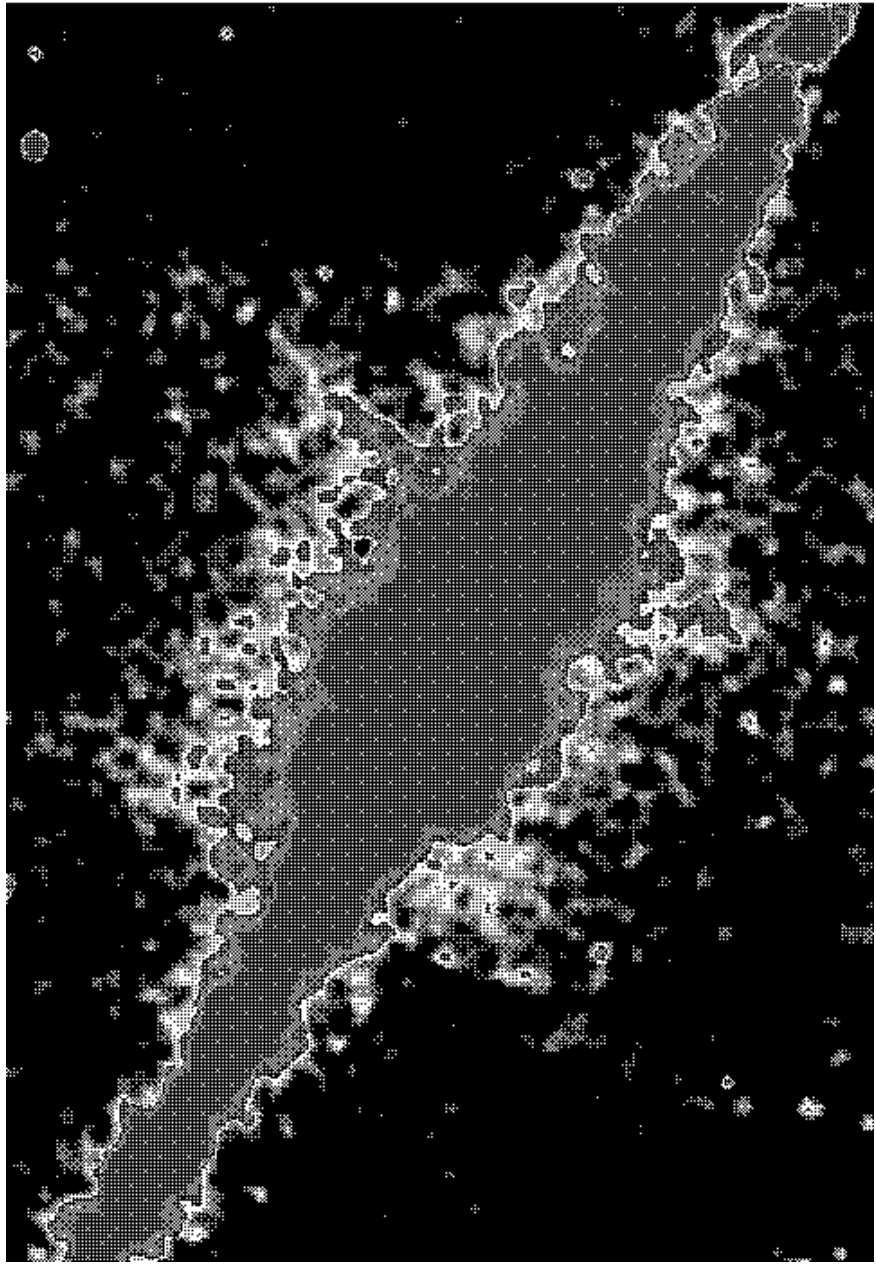}
\figcaption[LDMatthews.fig9_low_bw.ps]{Pseudocolor representation of the
\HA+[\NII] image of the inner regions of UGC~10043, after
smoothing with a Gaussian to $\sim6''$ resolution. The field shown is
roughly \am{1}{5} across. The emission along 
the midplane is ``burned in''
in order to emphasize the faint, vertically extended emission,
rising to $\sim\pm22''$ (3.5~kpc) above and below the midplane and 
forming a roughly biconical
structure. 
\protect\label{fig:HAsmooth}}
\suppressfloats
\end{figure}

\begin{figure}
\epsscale{1.1}
\plotone{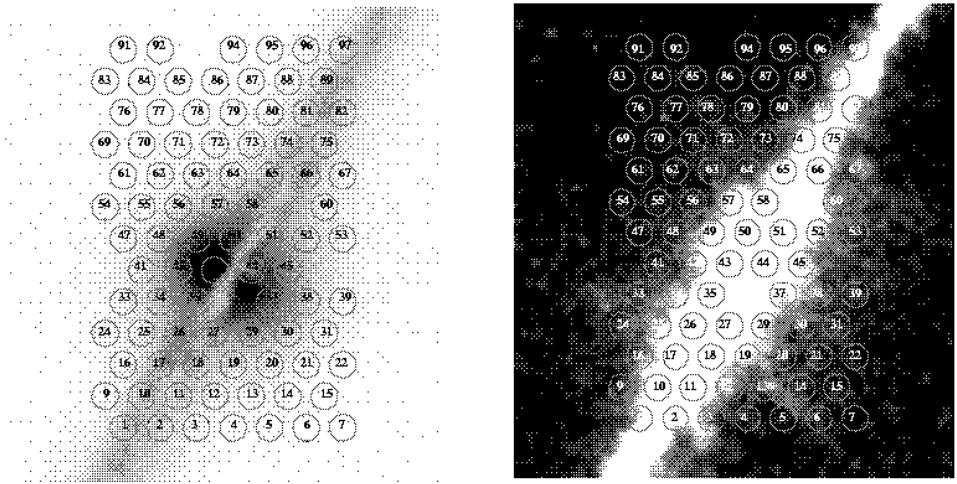}
\figcaption[LDMatthews.fig10_low_bw.ps]{Locations of the DensePak fibers 
overlaid on images of
UGC~10043: $R$-band (left) and a smoothed \HA+[\NII] image (right). 
The fibers have 3$''$
diameters and separations of \as{4}{15}. The fields shown are roughly $1'$ on
a side.
\protect\label{fig:denseoverimages}}
\suppressfloats
\end{figure}

\begin{figure}
\plotone{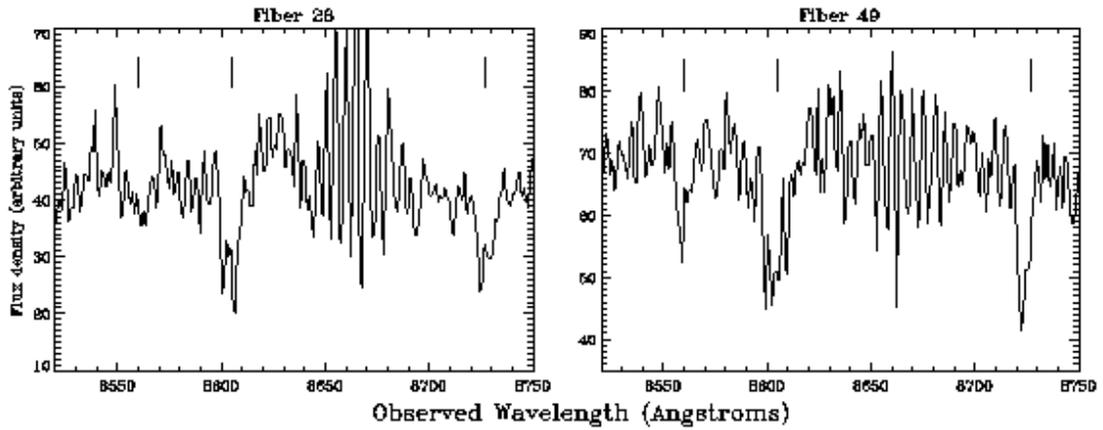}
\figcaption[LDMatthews.fig11.ps]{Two sample spectra of UGC~10043 
in the \CaII\ infrared triplet
wavelength
region, shown to illustrate examples of 
the quality of the data and the observed \CaII\ line shapes.  
The vertical lines denote the locations of the 
\CaII\ absorption features.
The apparent ``emission'' features are imperfectly subtracted night
sky lines. In a number of fibers, including the example on the left,
it appears the lines may be resolved into multiple velocity
components.
\protect\label{fig:CaIIsamples}}
\suppressfloats
\end{figure}

\begin{figure}
\epsscale{0.9}
\plotone{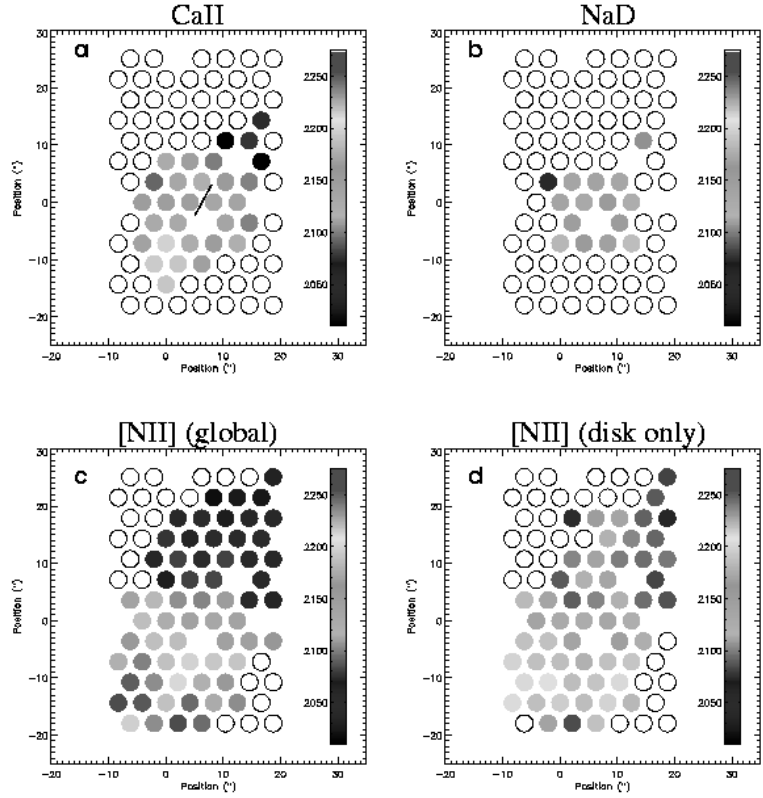}
\figcaption[LDMatthews.fig12_bw.ps]{Heliocentric velocity as a function of position in
UGC~10043, based on various spectral line 
measurements: (a)
cross-correlation analysis of the \CaII\ infrared triplet; (b)
mean value from Gaussian fits to the  two components of the Na~D
lines; (c) value from single-Gaussian fits of the stronger line of the
[\NII] doublet; (d) Component~A of two-line Gaussian decomposition of
the [\NII] line (see Section~\ref{ionkin}). The solid
line in panel (a) indicates roughly the location and orientation 
of the galaxy midplane and the galaxy center.
\protect\label{fig:bmontage}}
\suppressfloats
\end{figure}

\begin{figure}
\plotone{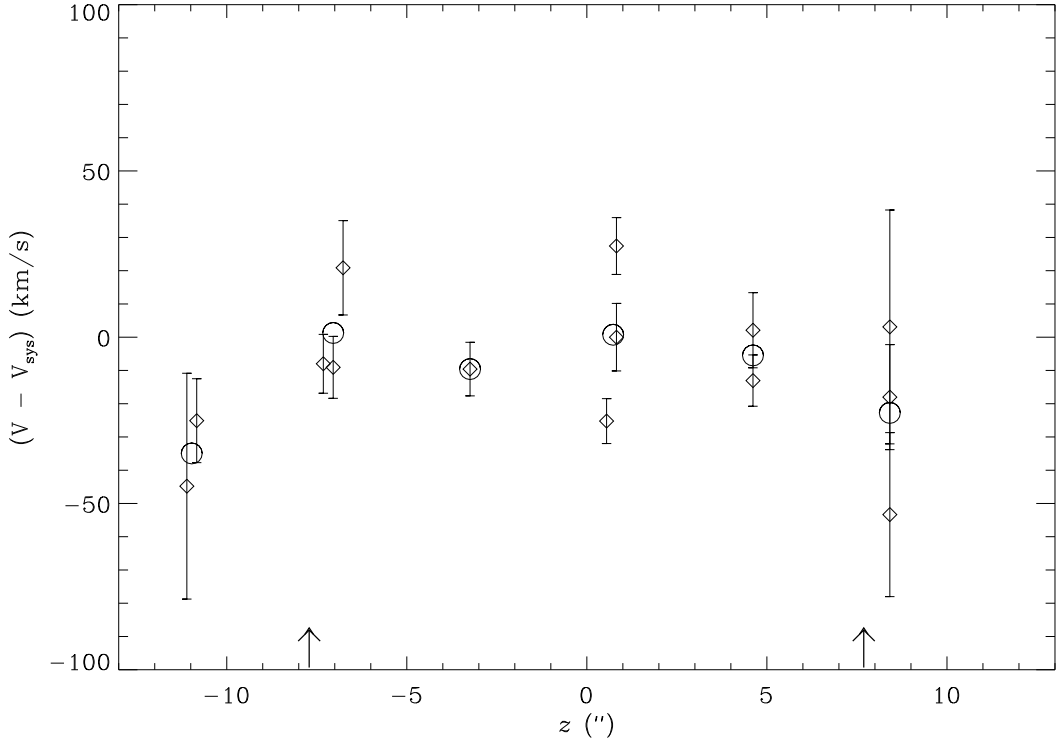}
\figcaption[LDMatthews.fig13.ps]{Minor axis rotation curve for 
UGC~10043 derived using
the \CaII\ velocities for all
fibers within $|r|<5''$ of the galaxy center where \CaII\ was
detected. Negative $z$ values correspond to the southwestern portion
of the galaxy. Data points at a given $z$ correspond to a range of $r$ 
values, hence their velocity spread reflects observed scatter as well
as contamination from the
underlying rotation of the disk along the major axis. The (unweighted) 
mean values at each
$z$  (representing the mean rotational speed along the minor
axis) are indicated by open circles. The arrows along the $x$-axis denote the locations
corresponding to the transition from prolate to oblate bulge isophotes (see
Section~\ref{bulge} and Figure~\ref{fig:bmontage}).
\protect\label{fig:CaIIminor}}
\end{figure}

\begin{figure}
\plotone{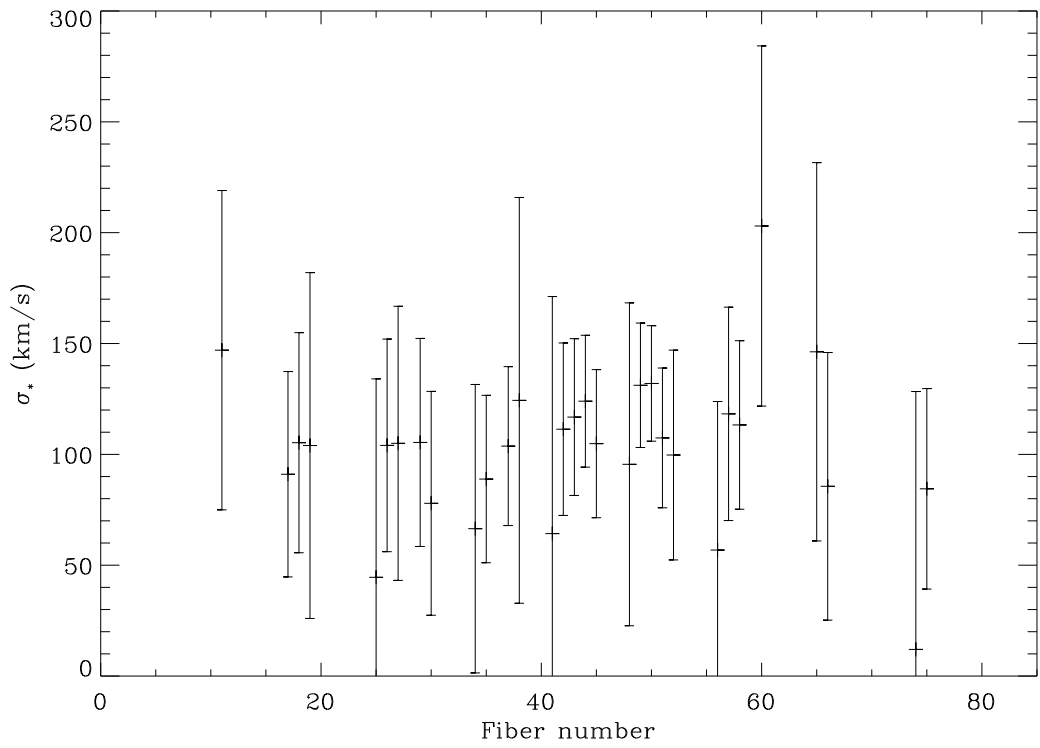}
\figcaption[LDMatthews.fig14.ps]{Stellar velocity dispersions in 
UGC~10043 inferred from the 
cross-correlation analysis of the \CaII\ infrared
triplet lines, as a function of fiber number. 
\protect\label{fig:veldisp}}
\end{figure}

\begin{figure}
\vspace{-1.5in}
\plotone{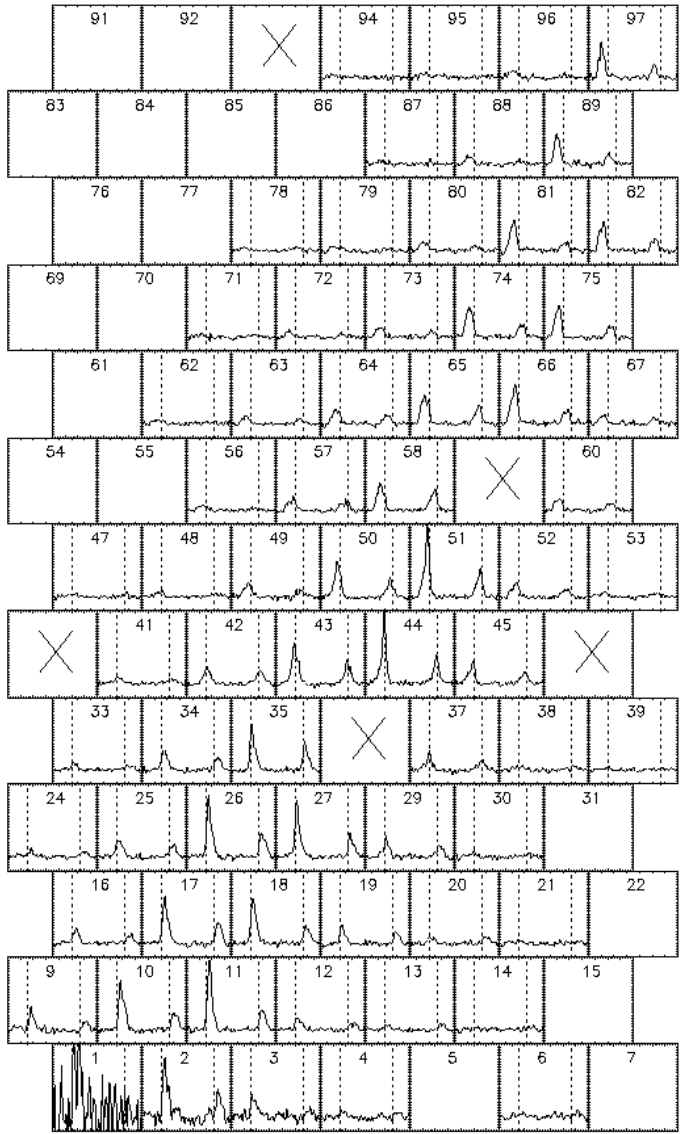}
\figcaption[LDMatthews.fig15_low.ps]{Schematic representation of 
the \HA\ and [\NII] emission lines as a function
of position in UGC~10043. North is on top, east on the left. Each
panel shows the DensePak spectrum in a single fiber, with the fiber
number indicated on each panel. Only fibers containing lines detection of
$\ge2.0\sigma$ across four or more pixels 
are shown. Due to the limited resolution of the plot,
many of the line peaks appear weaker than in the full-resolution
spectra. Fiber~1 has a
significantly lower throughput than the other array fibers,
hence the data are noisier.
The true physical scale of the array and the locations of 
the individual fibers on the galaxy are indicated in
Figure~\ref{fig:denseoverimages}. 
The spectra are plotted in the galaxy rest frame and
have the underlying stellar component subtracted
(see Section~\ref{ratios}). All plots have 
the same flux density scale (in arbitrary units), and show the wavelength 
interval 6555-6590\ang. The weaker line of the [\NII] doublet is
outside this interval.
Dotted lines indicate the rest wavelengths of the two emission lines. Empty
boxes indicate positions where no lines were detected (at $>2\sigma$); boxes
containing an `X' correspond to dead fibers.
\protect\label{fig:HAgrid}}
\end{figure}

\begin{figure}
\plotone{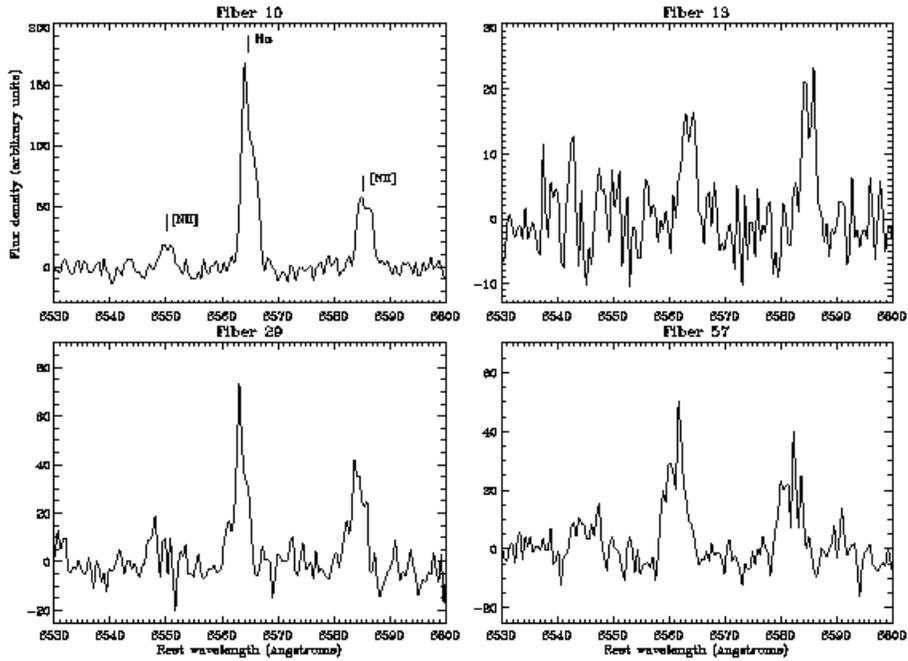}
\figcaption[LDMatthews.fig16_low.ps]{Four sample spectra of 
the \HA+[\NII] doublet wavelength
region. The spectra were shifted to the galaxy rest frame and have the 
underlying stellar component subtracted.
The line profiles are complex, and in each case appear resolved into two or more
velocity components. Fiber~13 is an example of a case where the 
[\NII] line is stronger than \HA. The fiber~57 spectrum shows an
example of the type of broad lines (FWHM$>$4\ang) that were seen at some
locations above and below the galaxy midplane.
\protect\label{fig:HAsamples}}
\end{figure}

\begin{figure}
\plotone{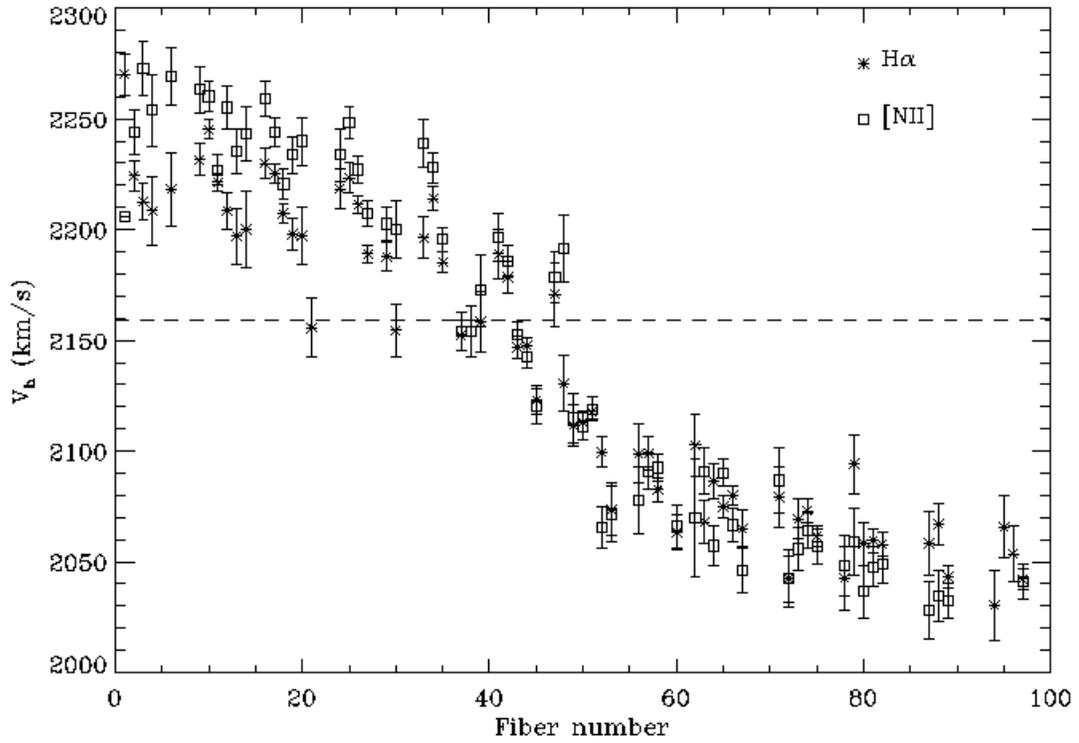}
\figcaption[LDMatthews.fig17.ps]{Heliocentric radial velocities 
determined for UGC~10043
using centroids determined from 
single-Gaussian fits to the \HA\ 
(asterisks) and [\NII] lines (squares). The systemic velocity is
indicated by the dashed line.
\protect\label{fig:HAvsNIIsingle}}
\end{figure}

In Figure~\ref{fig:bmontage}c we plot the [\NII] velocities as a
function of position in the galaxy, while in 
Figure~\ref{fig:HAvsNIIsingle} we plot the radial velocities derived from the
single-component fits as a function of fiber number for both the \HA\
and [\NII] lines. 
Surprisingly, for many
of the fibers plotted in Figure~\ref{fig:HAvsNIIsingle}, 
an offset of up to $\sim$60~\kms\ 
is seen between the inferred \HA\ and [\NII] velocity centroids. This
offset is
significantly larger than the expected velocity errors. Furthermore, these
velocity offsets cannot be the result of a wavelength calibration error, as both
lines of the [\NII] doublet show the same behavior, and their
wavelengths bracket that of \HA. We have verified that no
night sky or diffuse interstellar lines are contaminating the \HA\
measurements, and that this offset is not the result of residual contamination
from underlying stellar absorption in the \HA\ line (see
Section~\ref{ratios}). Moreover, 
both redward and blueward of the systemic velocity, the
offset between the \HA\ and [\NII] velocities 
is such that the [\NII] lines lie systematically further 
from $V_{\rm sys}$ than the \HA\ lines, suggesting the origin is likely to
be a real, physical effect.  Examining the
spatial distribution of the velocity offsets, we find that
for fibers lying near the disk midplane, the offsets are
no more than $\sim$10~\kms, the size of the typical error
bars. However, the offsets increase with increasing $z$-height, and
reach their most extreme values ($\sim$60~\kms) near the northeast and southwest
``corners'' of the hourglass-shaped distribution of ionized gas
highlighted in Figure~\ref{fig:HAsmooth}. 

Because a significant fraction of the emission line profiles in
UGC~10043 appear complex and could not be well-represented by only a
single Gaussian (Figure~\ref{fig:HAgrid} \& \ref{fig:HAsamples}),
we performed a second set of velocity measurements to these spectra
where we decomposed the lines using two Gaussian components. 
Once again, no constraints were imposed on the line
widths or relative wavelength separations. 
Based on these
two-Gaussian decompositions, we find a mean velocity dispersion for
both the
\HA\ and [\NII] line components of $\sim$0.6\ang\ ($\sim$25~\kms) 
after correction for instrumental broadening.

\begin{figure}
\epsscale{0.8}
\plotone{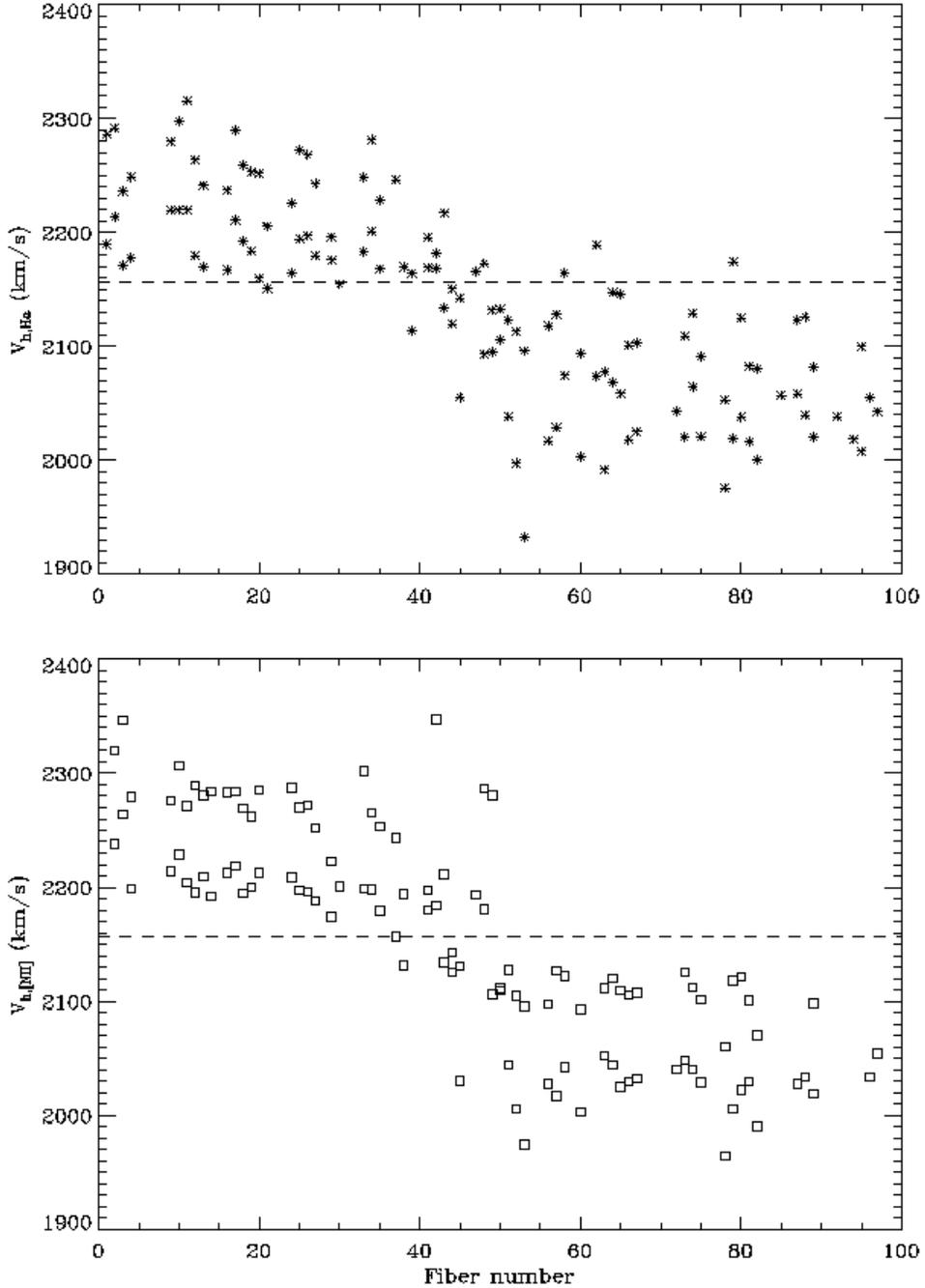}
\figcaption[LDMatthews.fig18.eps]{Heliocentric radial velocities as a 
function of fiber
number for the \HA\ lines (top)
and the [\NII] lines (bottom) determined for UGC~10043 using fits with
up to two Gaussian components. The heliocentric velocity is indicated
by the dashed line. The velocities from the
[\NII] line decompositions group into two rather distinct components
(referred to as A \& B in the text; see Section~\ref{ionkin}); 
the existence of distinct
components is less obvious in the \HA\ data. In addition, the
\HA\ decompositions reveal a series of points near the systemic
velocity that are not seen in the [\NII] decompositions.
\protect\label{fig:HAvsNIImulti}}
\end{figure}

In Figure~\ref{fig:HAvsNIImulti} we
have plotted 
the velocities derived from the two-component
decompositions of the \HA\ lines (top) and the [\NII] lines (bottom), both as a
function of fiber number.
It can be seen that the [\NII] lines
predominantly decompose into
two cleanly separated velocity components, with typical offsets of 
$\sim$70-80~\kms. The \HA\ components show a slightly larger spread of
velocities, and in spite of comparable signal-to-noise,
these lines do not cleanly decompose  into distinct groupings.
Part of this difference may be the result of
the underlying stellar absorption in the \HA\ line, which 
could introduce additional uncertainties in the \HA\ emission line
decompositions.
However, at a number of locations 
the \HA\ line decompositions show the presence
of a velocity component within $\pm$20~\kms\ of 
$V_{\rm sys}$, whereas no analogous component
emerges from the  [\NII] fits. With two exceptions, this occurs for
fibers either well above or below the midplane ($|z|>5''$).
Interestingly, roughly half of the fibers containing evidence of 
additional \HA\ emission near $V_{\rm sys}$ correspond
to the locations where the velocity offset 
$|V_{{\rm H}\alpha}-V_{\rm NII}|$ is observed to be the largest
based on the one-component Gaussian fits presented 
in Figure~\ref{fig:HAvsNIIsingle}.
This raises the possibility that the \HA\ lines observed at some locations may
actually be comprised of three or more distinct components, including one near
$V_{\rm sys}$ (which in turn would bias the one- and two-Gaussian
decompositions to lower velocities). However, the physical explanation
for such a component present only in \HA\ is unclear, and this possibility
will require investigation with additional high spectral resolution
observations. 

\begin{figure}
\plotone{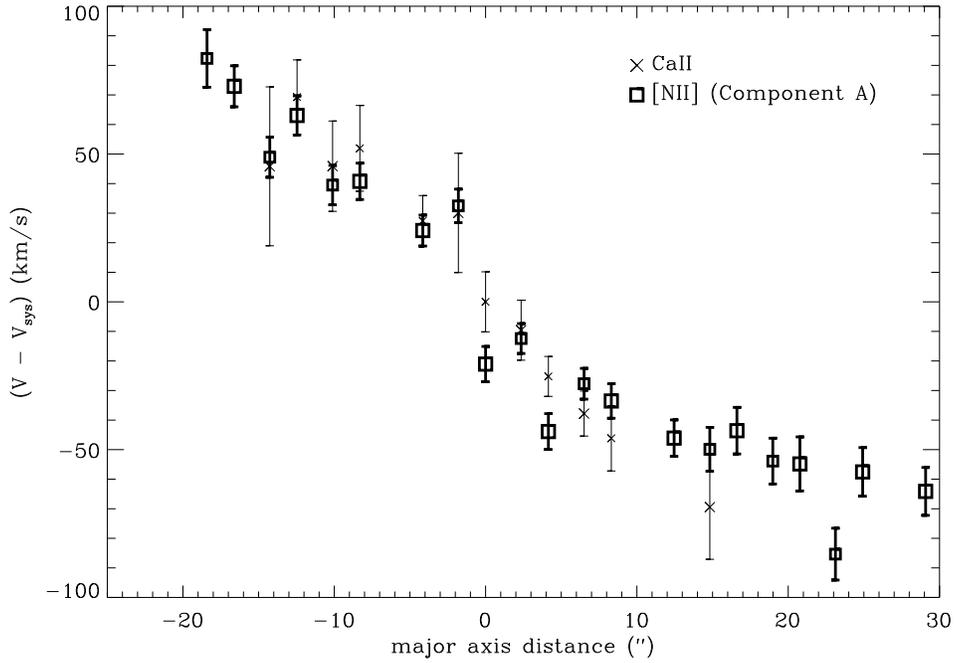}
\figcaption[LDMatthews.fig19.ps]{Rotational velocities along the 
major axis of UGC~10043 obtained
using Component~A of the [\NII] emission lines (bold squares;
see Section~\ref{ionkin}) and from the \CaII\ infrared triplet 
(crosses; see Section~\ref{CaIIkin}). Velocities from fibers  
10-17-26-35-43-50-58-65-74-81-89-97 are plotted as larger symbols and those from
fibers 2-11-18-27-44-51-66-75-82 as smaller symbols. The \CaII\
velocities are uncorrected for asymmetric drift, and none of the
velocities are corrected for line-of-sight
integration effects. See
Figure~\ref{fig:denseoverimages} for fiber locations.
\protect\label{fig:NIImajor}}
\end{figure}

In light of the trends described above, 
what are the physical implications of the emission line shapes
observed in UGC~10043? For the [\NII] lines, the interpretation
is more
straightforward than the seemingly more complex \HA\ lines.
In the case of the [\NII] lines,  the velocity
component nearer to the systemic velocity (hereafter Component~A)
follows the  pattern expected for the
rotation of the underlying disk. We plot the Component~A velocities
in Figures~\ref{fig:bmontage}d \& \ref{fig:NIImajor}. For comparison, in
Figure~\ref{fig:NIImajor} we have also
overplotted  the corresponding velocities derived from the \CaII\
lines in Section~\ref{CaIIkin}. Although the stars and the ionized gas
in UGC~10043 are not necessarily expected to share the same spatial
distribution, at most locations 
the velocities show good agreement, despite the neglect of asymmetric
drift corrections for the \CaII\ values.

The peak rotational velocities traced by the [\NII] lines are
$\sim$80~\kms\ and 85~\kms\ on the two sides of the galaxy,
respectively, and occur near the
last measured points at $r\approx20''$. These values 
are significantly less than the peak rotational velocities inferred from
global \HI\ observations ($V_{{\rm max}}\approx160$~\kms;
Section~\ref{radio}), implying the
rotation curve is still rising throughout the 
region sampled by the DensePak data. 

While Component~A of the [\NII] lines traces the underlying 
disk rotation, the second,
higher velocity component of the [\NII] lines (hereafter Component~B), 
presents the expected signature of 
a large-scale, cylindrically symmetric, biconical wind 
(e.g., Heckman et al. 1990). 
The existence of this wind was already suggested by the narrow-band
imaging observations presented in Section~\ref{narrowband}, and these
new emission line spectra provide firm kinematic confirmation of its existence.
Because UGC~10043 is seen so nearly edge-on, it is expected
that most lines-of-sight will intersect only a single
cone of the wind (see the schematic 
cartoon in Figure~19 of Heckman et al. 1990), and this is consistent with
the observed velocity structure of the lines. The line decompositions
also imply that we are primarily seeing emission only from the near
sides of the bicones.

Using a combination of the new imaging and spectroscopic data, 
it is possible to estimate some rough properties of the
galactic  wind of UGC~10043.
From Figure~\ref{fig:HAsmooth}, we estimate an opening angle for the
wind to be $\theta\sim 100^{\circ}$.  Assuming the bulk of the 
ionized gas entrained in the wind 
is outflowing tangentially along the surfaces of the bicones, the
outflow velocity can be expressed as 
$V_{\rm out}=\Delta V/[{\rm sin}(\theta/2)]$, where
$\Delta V$ is the observed velocity offset
between the wind material and the material
following the normal galactic rotation at a given position.
Taking $\Delta V\approx$80~\kms\ (Figure~\ref{fig:HAvsNIImulti}) yields 
$V_{\rm out}\gsim$104~\kms. Typical outflow velocities for
``superwind'' galaxies tend to lie in the range 200-1000~\kms\ (e.g., Heckman
2001), whereas the wind speed inferred for UGC~10043 is 
comparable to the inferred expansion velocity for the 
weak bipolar wind at the
Galactic Center (Bland-Hawthorn \& Cohen 2003) and to the 
high end of expansion velocities
measured for supershells in dwarf galaxies (e.g., Martin
1998). UGC~10043 thus appears to have one of the most feeble large-scale,
collimated winds presently known.
We further discuss the energy budget from massive stars
in UGC~10043 in Section~\ref{radio}, and we
speculate on a possible triggering mechanism for the 
wind in Section~\ref{Wakamatsu}.

\subsubsection{Minor Axis Kinematics of the Ionized 
Gas\protect\label{minorions}}
As discussed in Section~\ref{optmorph}, 
the presence of a dust lane along the minor
axis of UGC~10043 (which in turn, may be part of an inner polar ring
structure) hints at the presence of material in the bulge region whose angular
momentum is misaligned relative to 
that of the disk. Based on the DensePak absorption line spectroscopy we
have obtained, no evidence of orthogonal rotation
was detected in the stellar component of UGC~10043
(Section~\ref{CaIIkin}). An alternate possibility is 
that there may be orthogonally
rotating material present that is predominantly gaseous, as in the case of
minor axis dust lane ellipticals (e.g., Bertola 1987; Sage \& Galletta
1993) or the orthogonally-decoupled gas disk in the core of the Sa
spiral NGC~2855 (Corsini et al. 2002).

Unfortunately, the search for orthogonally-rotating ionized gas in UGC~10043 is
complicated by the large-scale galactic wind, whose
kinematic signatures already produce a complex imprint on the line
profile shapes. Nonetheless, such a component
might still manifest itself in the form of a systematic shift in the
intensity-weighted velocities along the direction of the minor axis.

\begin{figure}
\plotone{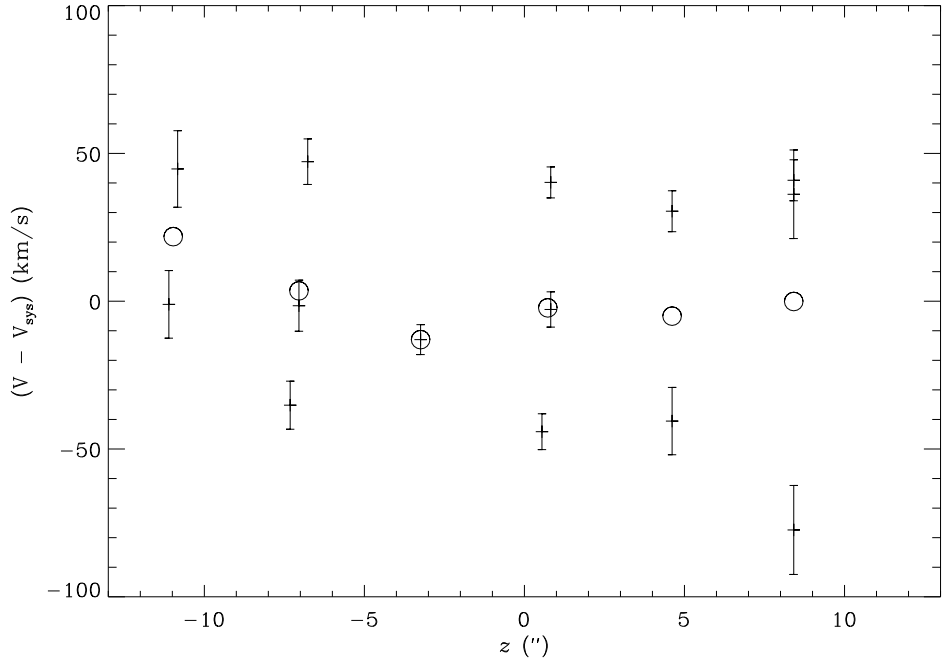}
\figcaption[LDMatthews.fig20.ps]{Minor axis rotation curve 
for UGC~10043 derived using
velocities from single-Gaussian fits to the
[\NII] lines. All fibers within
$|r|<5''$ 
and $|z|<12''$ of the galaxy center are plotted 
(41-48-56,42-49,35-43-50,44,29-37-45,30-38). Data points at
a given $z$ correspond to a range of $r$ 
values, hence their velocity spread primarily reflects the corresponding 
underlying rotation along the disk major axis. The (unweighted) 
mean values at each
$z$ (representing the mean rotational speed along the minor
axis) are indicated by open circles.\protect\label{fig:NIIvsHAminor}}
\end{figure}

In Figure~\ref{fig:NIIvsHAminor}
we investigate the kinematics of the ionized gas
along the apparent minor axis of UGC~10043 using the mean velocities derived
from single-Gaussian fits to [\NII] lines. 
As was the case with the stellar \CaII\ lines, no evidence for a 
significant 
minor axis rotational gradient is apparent 
($V_{\rm rot}\lsim$20~\kms); a similar 
result is seen in \HA. We have also examined the 
two-component line decompositions to search for rotational
signatures or other velocity peculiarities associated with the
locations of the minor axis dust lane and possible inner polar ring
structure, but no unambiguous evidence of orthogonal rotation is seen.
We are forced to conclude that although the UGC~10043 imaging data provide
tantalizing hints of
orthogonally-rotating material in this galaxy, 
if such a component is indeed present, its kinematic signatures are
subtle, and
our current spectroscopy does not have sufficient sensitivity and resolution
to verify its existence. 

\subsection{Emission Line Ratios\protect\label{ratios}}
Although the time-variable throughput of the DensePak fibers prohibits
accurate spectrophotometry with this instrument, it is still possible
to examine
the emission line intensity ratios ([\NII]/\HA) as a function of
position in the galaxy in order to probe possible
changes in temperature or ionization state.  Owing to the uncertainties inherent in
uniquely decomposing the emission lines into multiple velocity components, 
we do not attempt to determine
the line ratios for the individual velocity components from the
present data,
but concentrate only on the  {\it global} values.

Before quantifying the [\NII]/\HA\ line ratios, it is necessary to
account for stellar absorption underlying the interstellar
\HA\ emission lines. To correct for this
effect, we used synthetic galaxy absorption line spectra generated over the
wavelength range of our observations. These were produced using the
spectral synthesis code of Bruzual \& Charlot (2004).
The models assumed a Salpeter initial mass function (with lower and
upper mass cutoffs of $m_{L}=0.1M_{\odot}$ and $m_{u}=100M_{\odot}$,
respectively), solar metallicity,
zero dust reddening, an instantaneous burst of star formation, and an age
of 8~Gyr. However,
over the wavelength range of interest, and for intermediate 
age stellar populations, 
the results are  insensitive to metallicity and
burst duration.  We found models with 3\ang\ resolution to give a
good match to the observed widths of the galaxy lines. After appropriately 
redshifting the model template
spectra, we normalized the continuum levels between the models and
each spectrum, and then subtracted the suitably scaled template from the data 
to remove the stellar component. 

\begin{figure}
\plotone{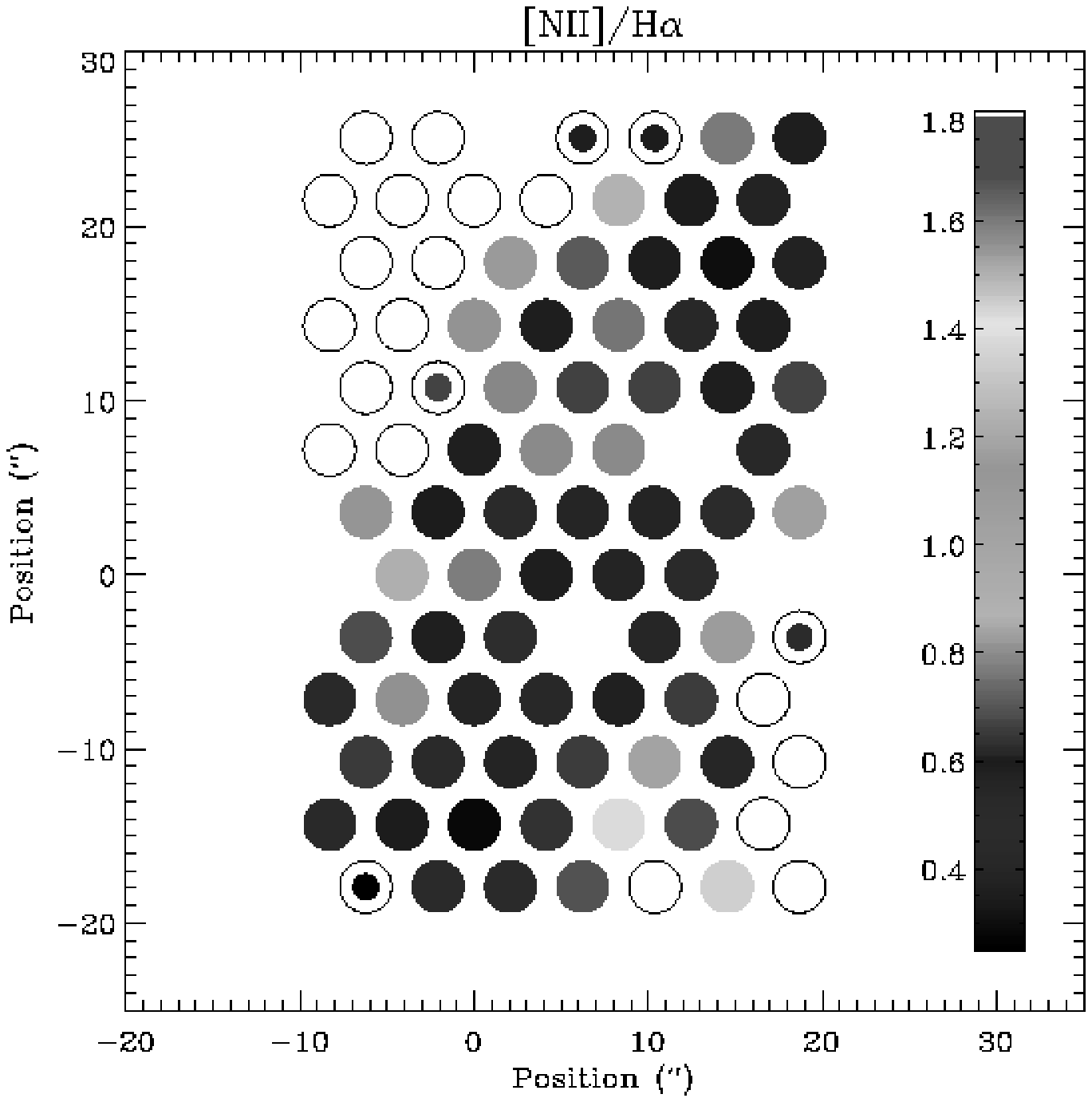}
\figcaption[LDMatthews.fig21_low_bw.ps]{The line ratio [\NII]/\HA\ as a 
function of position in
UGC~10043. All values have been corrected for the stellar
absorption underlying the \HA\ line. The smaller dots represent upper limits.
\protect\label{fig:linerat}}
\end{figure}

We plot the resulting [\NII]/\HA\ line ratios as a function of
spatial position in Figure~\ref{fig:linerat}. 
Along the disk of the galaxy, the observed [\NII]/\HA\ ratios
range from $\le$0.18 (an upper limit) to 0.80, with higher values 
concentrated towards the central regions.  
The mean value for the 23
fibers closest to the midplane ($|z|\lsim4''$) is [\NII]/\HA=0.43$\pm$0.14---very
typical of \HII\ region values (e.g., Miller \& Veilleux 2003).
Away from the midplane however  ($|z|>5''$),
there is a clear trend of increasing [\NII]/\HA\ ratios, including 8
fibers where
[\NII]/\HA$>$1.

Similar increases in the [\NII]/\HA\ ratio with increasing $z$-height have
now been observed in a number of edge-on spiral galaxies, both with and
without winds (e.g., Lehnert \& Heckman 1996; 
Rand 1998; Otte, Reynolds, \& Gallagher 2001; 
Miller \& Veilleux 2003). However, the physical origin of
this effect has been a long-standing controversy. In particular, it has
been argued that sources of ionizing heating in addition to photoionization
from hot stars (e.g., shocks or turbulent mixing layers) may be
required to explain the most extreme line ratios 
(see Miller \& Veilleux 2003 and references therein). For a
galaxy with a collimated wind, the generation of the wind involves the
expansion of a shell of hot gas into an
ambient medium; this in turn 
is expected to produce a shell of shocked gas  (e.g.,
Chevalier \& Clegg 1985). This could seemingly offer a natural explanation
for the $z$-dependent line ratios observed in UGC~10043. However, given the
relatively low inferred outflow speed for the wind, it is not clear
how effective such shock heating would be. In addition,
recent photoionization models of disk galaxies 
that take into account 
the multiphase nature of the ISM and the hardening of
the stellar radiation field with increasing $z$ can predict
[\NII]/\HA\ ratios as high as $\sim$1.5 
(Bland-Hawthorn, Freeman, \& Quinn 1997; Wood \& Mathis 2004). 
In UGC~10043, [\NII]/\HA$>$1.5
is observed only in one fiber (fiber 14, where [\NII]/\HA=1.7),
so the role of shocks versus photoionization in accounting for the
observed line ratios  remains unclear.
Ultimately,
observations of additional emission line diagnostics, coupled with
three-dimensional modelling, will be needed to better explore this
question.

\subsection{Detection of the Na~D Lines in UGC~10043}
After \HA, 
the strongest stellar absorption line expected
within the wavelength range covered by the 
DensePak \HA+[\NII] spectra is
the Na~D  doublet at $\lambda\lambda$5889.95,5895.92. This doublet is
prominent in the spectra of cool stars (particularly K and M spectral
classes). However, in many galaxies 
there can also be a significant interstellar contribution to
these lines (e.g., Heckman et al. 2000). 

The Bruzual \& Charlot models described in Section~\ref{ratios} predict that the
strength of
stellar Na~D absorption lines would make them essentially indistinguishable from
the noise in the present data. If the  mean stellar population is
somewhat younger than we have assumed,
the lines are expected to be even weaker, while for an older
population, any increase in line strength is negligible. 
Despite these predictions, we detect
the Na~D doublet in 15 fibers near the center of UGC~10043 
(Figure~\ref{fig:bmontage}b).
With one
exception, the corresponding fibers lie within the ``prolate'' region of the
inner bulge.
Interestingly, in all cases where Na~D was detected, these lines appear to be
not only deeper, but narrower than predicted by the (purely stellar)
Bruzual \& Charlot models. This strongly suggests these lines
are dominated by an {\it interstellar} 
contribution. Given our viewing geometry of UGC~10043
it is difficult to distinguish whether some of the Na-absorbing gas is associated
with the disk of UGC~10043, or whether it is all intrinsic
to the bulge region. The velocities of the Na~D lines (derived from
fitting a Gaussian+background to each component) do not help to
distinguish between these possibilities, since they match the
velocities of both those
of the \CaII\ lines (assumed to arise predominantly in the bulge;
Section~\ref{CaIIkin}) and Component~A of the 
[\NII] lines (associated with the disk; Section~\ref{ionkin}) to within errors.

\section{Radio and Far-Infrared Properties of UGC~10043\protect\label{radio}}
In order to estimate the energy budget from massive stars in
UGC~10043, as well as understand 
the conditions in its interstellar medium that
made it conducive to forming a detectable large-scale 
wind, it is of interest to examine the
radio and far-infrared properties of the galaxy.

UGC~10043 was observed in the \HI\ 21-cm line by Lewis et al. (1985) and
Freudling, Haynes, \& Giovanelli (1988), both using
single-dish telescopes. The spectrum of Freudling et al. was
subsequently reanalyzed by Giovanelli, Avera, \& Karachentsev (1997). 
UGC~10043 exhibits a classic double-horned \HI\
profile characteristic of a rotating disk. The Lewis et al. \HI\ spectrum
appears rather asymmetric, but this may be the result of a small pointing
error, as the Freudling et al. spectrum shows 
only a slight asymmetry in the peak flux of the
two horns. 
Both edges of the \HI\ profile are steep and 
have approximately equal slopes, showing no evidence for any recent
perturbations of the 
outer \HI\ disk. However, the spectrum shows an excess ``bump''  near
the systemic velocity (roughly 100~\kms\ wide and $\sim3\times$ the rms
noise), hinting there could be anomalous \HI\ 
gas in
the inner regions of the galaxy.

The maximum disk rotational velocity for a galaxy can be estimated as
$V_{{\rm max}}\approx\frac{1}{2}(W_{20} -
W_{{\rm rand}})$, where $W_{20}$ is the global \HI\ linewidth measured at
20\% peak maximum and $W_{{\rm rand}}\approx$20~km s$^{-1}$ accounts for the
typical random
component of the gas motions. For UGC~10043 this yields
$V_{{\rm max}}\approx$160~\kms\ (see Table~1). 
The total dynamical mass internal to the
last measured point may then be
estimated as
$M_{{\rm dyn}}=2.326\times10^{5}r_{{\rm max}}V^{2}_{{\rm max}} = 
1.0\times10^{11}~M_{\odot}$, 
where we have assumed
$r_{{\rm max}}\approx1.25R_{25.5}$=17~kpc.

Using the integrated \HI\ flux from Giovanelli et al. (1997), and
assuming the gas is optically thin, the \HI\
mass of UGC~10043 is \mhi$=5.9\times10^{9}~M_{\odot}$, and
\mhi$/L_{B}\sim$0.66 ($M_{\odot}/L_{\odot}$).
While the distribution of the \HI\ gas
in UGC~10043 is unknown, the shape of the global line profile suggests
gas distributed throughout an extended disk, and 
both the \HI\ mass and the fractional \HI\
content suggest the galaxy has ample raw fuel for
star formation and abundant gas-rich material which could become entrained in a
wind-driven outflow.

UGC~10043 was detected by {\it IRAS} in all four of its
bands. According to the NED database, the measured
60$\mu$m and 100$\mu$m fluxes are 1.164~Jy$\pm$5\% and
3.453~Jy$\pm$7\%, respectively. Using the definition of Helou et
al. (1988), this yields a far-infrared luminosity 
$L_{{\rm FIR}}=1.1\times10^{43}$ ergs~s$^{-1}$,  integrated between
40-120$\mu$m (see Table~1). 
This is roughly an order of
magnitude lower than the canonical values 
found by Lehnert \& Heckman (1996) to typify galaxies with
superwinds ($L_{{\rm FIR}}\gsim10^{44}$ ergs~s$^{-1}$). 
The ratio $S_{60\mu{\rm m}}/S_{100\mu{\rm m}}$=0.34 for UGC~10043
is somewhat smaller than the minimum values found by Lehnert \&
Heckman (1996)
in galaxies exhibiting
superwinds ($S_{60\mu{\rm m}}/S_{100\mu{\rm m}}\gsim$0.5), and also
falls below the cutoff for so-called ``infrared warm''
galaxies ($S_{60\mu{\rm
m}}/S_{100\mu{\rm m}}>$0.4; e.g., Lehnert \& Heckman 1996). In a
global sense, UGC~10043
therefore does not come close to classifying as a starburst galaxy.
Indeed, compared with the mean physical parameters for galaxies of
various Hubble types tabulated by Roberts \& Haynes
(1994), UGC~10043 is quite typical of normal Sbc spirals in terms of
its \HI\ mass, dynamical mass, and far-infrared luminosity, while
its optical luminosity falls somewhat below normal
(by roughly a factor of two)
giving it a \dark\ ratio more typical of giant low surface brightness
spirals
(e.g., Matthews, van Driel, \& Monnier Ragaigne 2001). 

The 20-cm radio continuum flux of UGC~10043 obtained by the NRAO VLA
Sky Survey (NVSS) 
is $F_{\rm cont,20}$=10.6$\pm$0.9~mJy (Condon et
al. 1998), making the galaxy consistent with the
radio-far-infrared correlation (Condon 1992). Hummel, Beck, \& Dettmar
(1991) also detected UGC~10043 at 6-cm, measuring
$F_{\rm cont,6}$=4.5$\pm$0.1~mJy. Together these two measurements imply a
steep spectral slope dominated by synchrotron emission (see Condon
1992). The detected radio continuum emission at both frequencies 
is confined
to the inner disk/bulge region of the galaxy. In the NVSS data, the continuum
emission is marginally resolved along the $r$-direction,
implying a deconvolved size for the emitting region of $\sim32''$
FWHM. From the 6-cm observations, Hummel et
al. (1991) also report a similar $r$-extent ($30''$); this is 
comparable to the angular extent of the wind-emitting region
(e.g., Figure~\ref{fig:wiynimages}).

The confinement of the bulk of the radio continuum emission to the inner regions
of UGC~10043 is consistent with other evidence for
centrally-concentrated star formation, including
candidates for dark clouds
across the inner 2-3~kpc seen in the optical images
(see Figure~\ref{fig:wiynimages}), 
as well as the mid-ultraviolet imaging data of
Windhorst et al. (2002), which when smoothed to increase
signal-to-noise, show the inner 15$''$ of the bulge region to be the brightest
portion of UGC~10043 in the mid-ultraviolet. 
Note that whereas some
bulges are detected in the
{\it far}-ultraviolet due to UV light from old,
low-mass stars, the {\it mid}-ultraviolet (center at 3000\ang) used by Windhorst
et al. primarily traces young (age
$\lsim$1~Gyr) stellar populations. Furthermore, that this UV emission is
associated with the bulge rather than the inner disk is 
suggested by its elongated $z$-distribution. We have verified that
this flux does not result from the red leak of the $F300W$ (mid-UV) filter (see
Biretta et al. 2000) by
folding our Bruzual \& Charlot model spectrum (Section~\ref{ratios}) through
the filter's WFPC2 response function via the STSDAS\footnote{STSDAS was
developed at the Space Telescope Science Institute. STScI is operated
bythe Association of Universities for Research in Astronomy, Inc. for
NASA.} SYNPHOT package. We find that only $\sim$1\% of the $F300W$
flux can be due to red leak.
Together these findings provide evidence for enhanced nuclear star formation
in UGC~10043, while at the same time
underscoring that the outer UGC~10043 disk appears of optically low surface
brightness owing primarily to a lack of stars rather than to heavy
dust obscuration.

Using the radio continuum and far-infrared luminosities, together with
the \HA\ luminosity from Section~\ref{phot}, we can obtain three 
estimates of the current total star formation rate (SFR) in
UGC~10043. To correct the \HA\ luminosity for extinction, we have
adopted the mean correction
for Sbc spirals found by Kewley et al. (2002) based on measurements of the
Balmer decrement ($\times$2.6). The
formulae of Kennicutt (1998) (for the FIR and \HA) and Bressan, Silva,
\& Granato (2002) (for the radio) then yield the following values
(assuming a solar abundance and a Salpeter initial
mass function):\footnote{The FIR estimate assumes
$L_{\rm IR}\approx 1.75L_{\rm FIR}$ (Kewley et al. 2002). Although
UGC~10043 adheres to the radio-far infrared flux correlation, the SFRs
inferred from the two tracers differ, largely owing to 
uncertainties in the assumptions regarding the contributions of 
the thermal versus
non-thermal component to the 20-cm flux (see Condon 1992; Bressan et
al. 2002).} (SFR)$_{20cm}=$1.06$\pm$0.08~$M_{\odot}$~yr$^{-1}$,
(SFR)$_{FIR}$=0.87$\pm$0.03$M_{\odot}$~yr$^{-1}$, and
(SFR)$_{H\alpha}\approx0.34\pm0.09M_{\odot}$~yr$^{-1}$. Here the uncertainties
account for the flux measurement errors only.
Interestingly, the \HA\ estimate value is nearly a factor of three
lower than the other two. This discrepancy is larger than expected solely
from the uncertainty in the correction for internal
extinction, and would become even more significant if we applied
extinction corrections appropriate for an Sd disk (see Kewley et al. 2002).
This provides further indirect evidence for 
optically-obscured star formation in the
nuclear regions of the galaxy.

The SFRs implied by the FIR
and radio continuum fluxes are very typical of the globally averaged 
SFRs for Sbc spirals 
(e.g., Kewley et al. 2002),
but are significantly lower than those that typify  starburst galaxies 
with superwinds (e.g., Heckman et al. 1990; 
Lehnert \& Heckman 1996). Nonetheless,
as emphasized by Heckman (2001), the criterion
for driving a large-scale wind depends far more critically on the SFR {\sl
per unit area} rather than on the globally averaged SFR.
Heckman (2001) quotes a SFR per unit area of $\ge 0.10
M_{\odot}$~yr$^{-1}$~kpc$^{-2}$ as the rule-of-thumb for powering a
superwind. Veilleux (2004) points out that this value is
conservatively high,
since the production of winds also depends on criteria such as the ages of the
stellar energy source and 
the geometry of the ISM. Nonetheless, taking this as a characteristic
value, we find that to reach this threshold in UGC~10043 requires that
nearly all of the star formation  
must be concentrated in a relatively small region
($d\lsim22''$, or $\lsim$3.6~kpc) near
the center of the galaxy. This size is comparable in extent to other 
star formation tracers, including
20-cm radio continuum, the brightest mid-ultraviolet emission, and the
brightest \HA\ emission. This size scale also is consistent with the
geometry of the wind, as a more extended disk or annulus containing
vigorous star formation would tend to
produce a thickened disk of ionized gas rather than an organized,
biconical wind. As discussed by Elmegreen (1999), intense star
formation within a bulge or nuclear disk may have an increased ability to
produce disk blowout compared with  star formation in other parts of the galaxy, 
since the  higher gas velocity dispersions
in these regions tend to decrease the self-regulation of the star formation.

\section{Speculations on the Formation History of 
UGC~10043\protect\label{discussion}}
Overall, the structural and kinematic complexities of UGC~10043,
including its peculiar, vertically-elongated bulge and
the presence of a large-scale wind,  suggest
a complex evolutionary history, most likely  
requiring a major ``second event'' (accretion, infall, and/or
merging). While these types of processes  are now believed to
play a general  role in galaxy formation,
UGC~10043 appears to be part of a special group of disk+bulge systems
(that includes the polar ring/disk galaxies and orthogonally-decoupled
bulge galaxies) whose formation
requires rather specific combinations of conditions (e.g., 
environment, merger geometry, impact
velocity, gas cross-section, dark halo concentration) 
to account for both the production 
and maintenance of their current
structures. As such, these galaxies uniquely preserve and showcase important
clues to their formation histories. And since
the parameter space for producing such systems is
far more limited than in the more general case, studies of galaxies like these can
help to lift some of the degeneracies in our understanding 
of how a variety of internal and external factors  influence
the present-day structures and
compositions of  galaxies. In this context,  the
unique properties of UGC~10043 make it a particularly
important laboratory  for testing various 
classes of evolutionary models. 
Drawing on
relevant theoretical modelling and discussions from the literature,
we now sketch some 
possible 
formation scenarios for UGC~10043. We outline the strengths and
weaknesses of each scenario in hopes of motivating additional
observations and numerical modelling  to further test and
refine these ideas. 

\subsubsection{Accretion of a Disk around a Naked Spheroid}
There is no obvious mechanism by which a disk galaxy with a
vertically-elongated bulge
could form either by collapse of a
single protogalactic cloud, or by secular evolution of an existing
disk (e.g., through development of a bar instability). For these reasons,
Bertola \& Corsini (2002) have suggested that such galaxies 
are likely to represent cases where disks formed around pre-existing, 
naked spheroids (see also Binney \& May
1996). In this picture, the decoupled, inner bulge cores seen in
several such systems would be by-products of the
disk-building event (see also Arnaboldi et al. 1993).

Although we do not yet know if it has a decoupled inner core,
UGC~10043 offers an important constraint on this type of
model, since its stellar disk is thinner, more regular, and more extended
than the disks of any of the other spirals in which vertically elongated bulges
have been identified (compare
Figure~\ref{fig:wiynimages} with the images
in e.g., Sarzi et al. 2000; Reshetnikov et al. 2002). 
The small scale height of the UGC~10043 stellar
disk and the relatively long characteristic dynamical time scales of its outer
regions imply it is a
dynamically old entity that has not undergone any dramatic heating or
perturbations within at least the last few Gyrs (e.g., Reshetnikov \& Combes
1997). This is also suggested by the observed vertical
and radial
color gradients, which are likely to arise from
dynamical processes acting over a significant fraction of a Hubble
time.  

The idea of ``dressing a naked spheroid'' would seem to account 
nicely for certain properties of UGC~10043, including 
the apparent disjointedness of bulge and disk 
(Section~\ref{optmorph}), and the
similarities of its bulge to an elliptical galaxy
(Sections~\ref{bulgedecom}, \ref{faberjackson}). In most galaxies,
disk-building is
also expected to augment an existing bulge (e.g., Kormendy 1993) and
may account for the observed bulge color gradient
(Section~\ref{cmap}), or perhaps even its triaxial nature. 
It can be considered a more extreme version of the mechanism believed
responsible for the formation of dust lane ellipticals (e.g.,
Bertola 1987; Sage \& Galletta 1993; Oosterloo et al. 2002). 
Indeed, this type of model has been
invoked to explain the origin of the Centaurus~A system
(e.g., Malin, Quinn, \& Graham 1983). However, in stark contrast with
UGC~10043, the dusty, minor-axis disk of Centaurus~A
is dramatically warped and twisted, presumably the  result of
violent shearing and subsequent precession of the accreted disk (Sparke 1996).

A modified version of the above scenario
would be the idea of disk growth around
a spheroid having a smaller, pre-existing, gas-poor disk. Variations of this 
latter picture have been
invoked as a means 
to account for the origin of  NGC~5084
(Zeilinger et al. 1992; Carignan et al. 1997),
NGC~4550 (Rix et al. 1992), and other examples of S0-like galaxies
possessing gas and/or
extended disks
(Bertola, Buson, \& Zeilinger 1992; Reshetnikov \& Sotnikova 1997;
Kannappan, Jansen, \& Barton
2004). 

Unfortunately, for the case of UGC~10043,  this picture seems to be faced
with some challenges.
First, the material accreted onto either a naked spheroid or an S0-like
galaxy would have to be
predominantly gaseous, since any stellar constituent would be 
dissipationless and unable
to cool into a thin layer. In addition, the gas would need to be located
predominantly outside the stellar disk of the donor galaxy in order
for it to be readily tidally stripped. Numerical simulations
show that gas fractions of up to 
$\sim$40\% may be readily accreted from a donor provided that the gas
is sufficiently extended
(Bournaud \& Combes 2003).  However,
the neutral gas content of UGC~10043
(\mhi$=5.9\times10^{9} M_{\odot}$) is typical of the  gas
contents of giant spiral
galaxies; this implies that to produce a galaxy as
gas-rich as UGC~10043 would require a rather gas-rich predecessor
(\mhi$\gsim 5\times10^{10}M_{\odot}$ to provide for both the current ISM and
stellar content), with a weak enough potential
such that its gas could be readily stripped by a galaxy of UGC~10043's
mass ($M_{\rm tot}\sim10^{11}M_{\odot}$). The giant low surface brightness
spirals might be a suitable class of donor, but these are relatively
rare, at least in the
local universe (e.g., Matthews et al. 2001 and references therein).

Another challenge for the accretion scenario is that
numerical simulations show that only rings or annuli, not 
complete disks, are directly formed via
accretion (e.g., Rix \& Katz 1991;
Reshetnikov \& Sotnikova 1997; Bournaud \& Combes 2003). Therefore
subsequent perturbations and viscous evolution 
would be required to produce radial spreading into a bona fide disk. 
This process is expected to take a minimum of several
orbital periods (i.e., at least several Gyr; Rix \& Katz
1991)---although this is not necessarily problematic, given
other dynamical arguments for an old age for the UGC~10043 disk. In addition,
Rix \& Katz (1991) have shown that as an accreted gas disk grows radially, it
will also decrease in
scale height as it transitions from a pressure-supported to a
rotationally-supported regime. This may offer a natural explanation
for the observed radial color gradients in UGC~10043.
However, in order to produce the thin
{\it stellar} disk observed in UGC~10043 would seem to demand that significant
cooling of the gas disk took place before significant star
formation ensued. In addition, in order to avoid strong
twisting and differential precession over time,  the accreted disk 
would likely need to be self-gravitating (Sparke 1986).
Numerical checks on whether the expected timescales
for these various processes could reproduce the observed UGC~10043 disk would
be valuable.

An alternate version of the accretion picture is 
that the UGC~10043 disk was built from multiple dwarf
galaxies or subgalactic clumps, much as in the standard
hierarchical build-up paradigm for normal disk galaxies. 
This may offer a natural explanation for the large
observed radial color gradient in UGC~10043 (see Section~\ref{cmap})
as well as its tilted inner disk and
outer warp (see Sections~\ref{optmorph}, \ref{cmap}).
Nonetheless, the idea of hierarchical build-up has at least one serious
drawback for the case of UGC~10043---namely, how to  build such a thin disk
without excessively heating it (e.g., T\'oth \& Ostriker 1992). 

Some hope is offered by the
recent numerical simulations of Huang \& Carlberg (1997) and
Val\'azquez \& White (1999), which have shown 
that while a merging galaxy on a prograde orbit can cause
substantial heating and damage to a disk, an interloper on a retrograde orbit 
will cause far less damage (owing to the weaker coupling between the
infalling material and motion in the existing disk), and instead 
induce a coherent tilt. 
Heating effects are further lessened if the accreted satellites
have small relative masses, and accretion primarily occurs onto the
dark halo. However, the accretion of a number of 
dwarfs of subgalactic clumps
would be needed to build a disk as massive as that of UGC~10043, and it is
unlikely that every infall event would have occurred on a retrograde
orbit or onto the halo. In
addition, if the potentials of 
``naked spheroids'' contain two
preferred planes (one along the short and one along the long axis),
and if this build-up mechanism were effective,
we might
expect vertically elongated bulges to be observed in a much more significant
fraction of present-day Sb-Sc spirals. 

\subsubsection{Capture}
A second possibility is effectively the inverse of the above accretion
scenario---namely
that the bulge of UGC~10043 may be a secondary structure, acquired
well after the disk had formed and settled. The culprit might be 
a small, gas-poor elliptical
galaxy plunging in at a very low impact velocity (less than a few tens of \kms)
along its pole. This is analogous to the mechanism proposed for the formation
of collisional ring galaxies, including the bulge-dominated
collisional ring galaxies known as Hoag-type objects (Appleton \&
Struck-Marcell 1996 and references therein).

Interestingly, numerous pure disk (Sd) galaxies with properties nearly identical
to the disk of UGC~10043 are known, and many examples
can be found in the Flat Galaxy Catalogue of Karachentsev,
Karachentseva, \& Parnovsky (1993). One example is UGC~825
(Matthews \& van Driel 2000; see their Figure~7). UGC~825 is strikingly
similar to the UGC~10043 disk not only morphologically (having a
``needle-like'' disk with a dusty center), but also
in terms of  surface
brightness, luminosity, linear size, rotational velocity, and \HI\
content. 
Small, compact ellipticals with luminosities
similar to the UGC~10043 bulge are
also known, 
although these seem to be relatively rare in the local universe 
(e.g., Wirth \& Gallagher 1984; Nieto \& Prugniel 1987).

That a pure disk galaxy might be able to ``capture'' a
bulge has the qualitative appeal of accounting for the
appearance of UGC~10043 as a disk galaxy whose bulge has been ``inserted''
post factum (Figure~\ref{fig:unsharp}); it  could also simultaneously account
for several of the unusual properties of the bulge (large vertical
extent relative to the disk, triaxial shape, lack of evidence for
significant rotation) that imply it could not have formed secularly.
Nonetheless, it remains highly questionable
whether this process could work at all in practice, as it would seem to
require an extraordinary set of circumstances, including a gas-rich
disk and a small, compact elliptical present in the same region of space, both
on suitable orbits, 
and possessing an
extremely low relative velocity, rarely observed in real galaxy groups.
Initially,
the victimized disk would be expected to develop a 
bowl-like shape and ring structures in the stars and gas---neither of
which is evident in UGC~10043---although 
for a collision of roughly equal-mass galaxies,
these signatures are expected to diminish significantly after $\sim10^{8}$~years.
(Gerber, Lamb, \& Balsara 1996).

Since 
the bulge and disk of  UGC~10043 share the same radial systemic velocity,
one further requirement for a ``capture'' scenario is that
the bulge cannot simply be caught in the act of ``passing
through''. 
However, Appleton \& Struck-Marcell (1996) propose that an 
intruder might be strongly decelerated via
dynamical friction from a massive halo in the target galaxy.  Very
thin disks such as UGC~825 are expected to possess massive 
dark matter halos
(e.g., Zasov, Makarov, \& Mikhailova 1991; 
Gerritsen \& de Blok 1999), although numerical simulations are needed
to test whether they could
supply sufficient dynamical friction to effectively halt an intruder.

\subsubsection{Origin via a Major Merger}
Finally, let us consider the possibility that both the bulge and disk
of UGC~10043 formed through a single event involving the
merger of two moderate mass
disk galaxies. This model has been explored by Bekki
(1998) as a means of accounting for the origin of
polar ring galaxies whose rings are too massive, too gas-rich, and too
radially extensive to be readily
explained via mass transfer or capture of a dwarf satellite (see also
Iodice et al. 2002;
Bournaud \& Combes 2003). In this picture, a gas-rich ``intruder''
plummets through its ``victim'', 
leading to the formation
of a central S0-like component, a stellar halo, 
and an orthogonally rotating ring or annulus
of gas and stars via a combination of 
violent relaxation and dissipative processes. To
produce a large ($r>$10 kpc) disk also requires the victim to be a
gas-rich system (Bournaud \& Combes 2003).

At least qualitatively,
this scenario appears able to account for certain characteristics  of
UGC~10043. First, this type of event is expected to drive gas to small radii,
where it could trigger a nuclear
starburst. 
In addition, it offers a way of accounting for some of the unusual
properties of the bulge (vertically elongated isophotes; lack of
evidence for major axis rotation). 
Through dissipative cooling, it also may be able to produce a 
gaseous structure with very small scale height.
Unfortunately, like the scenarios described above, this one too seems
to meet with some 
shortfalls and the uncomfortable requirement of rather specialized
initial conditions. As discussed by Bournaud \& Combes (2003), to produce a
very massive polar disk requires that the victim itself be a very
massive, gas-rich disk (with \mhi\ roughly twice that of the
resultant system)
and quite stellar-poor, so as to not produce an observable 
stellar halo. Thus the victim would need to be something resembling a
giant low surface brightness galaxy. An additional constraint is
similar to that cited for the ``capture'' scenario---namely that to form a
disk rather than a ring, the impact
speed of the intruder would need to be less than 30-40~\kms---a rare condition in
typical galaxy groups. Neither of these conditions rules out the
merger picture, but it does imply that galaxies like UGC~10043 should
be quite rare. Another concern is whether such a violent event could
reproduce the complex, triaxial bulge structure of UGC~10043 and the
significant color differences between bulge and disk.
Once again, numerical work is needed to further explore
this possibility.

\subsection{Origin of the Enhanced Nuclear Star Formation and 
Galactic Wind\protect\label{Wakamatsu}}
Independent of the details of how UGC~10043 was assembled, another
question surrounding this galaxy concerns how it has evolved since
formation, and what has led to the
production of its large-scale wind
and centrally-concentrated star formation. Although most of the types of
formation mechanisms described above are predicted to drive gas to the
inner regions of the galaxy (e.g., Reshetnikov \& Sotnikova 1997;
Bournaud \& Combes 2003),
the gas consumption rate
required to feed the wind of UGC~10043 
($>1M_{\odot}$ yr$^{-1}$ within the inner
few kpc of the galaxy) suggests that the
wind is likely to be a transient and recently-triggered
phenomenon. Since the dynamically settled appearance of the disk of
UGC~10043 suggests it is a relatively old structure, it
is therefore not clear that the event that formed the UGC~10043
system is the same one
that instigated the current  nuclear star formation and wind.

One possible solution to  this 
puzzle may be the mechanism proposed by Wakamatsu
(1993). Wakamatsu suggested that if the central component of a polar ring
galaxy is sufficiently dense, its
oval potential will severely disturb
the orbits of the gas at small and intermediate galactocentric radii
in the disk, leading to the production of shock waves. 
He argues that repeated shock waves occurring as the disk rotates 
would cause loss
of orbital energy and angular momentum, and in turn drive gas to the
nuclear regions of the galaxy, much in the same manner as an inner bar. 
While we have shown that UGC~10043 is not a true
polar ring (or polar disk) galaxy, its vertically-elongated
bulge may nonetheless present a potential similar to the one envisioned by
Wakamatsu (1993). Moreover, to match observations of  polar ring galaxies
such as NGC~4650A, Wakamatsu assumes that the polar ``ring''
in his model galaxies is actually a disk-like entity, making
the overall situation seemingly rather analogous to the UGC~10043 system.
Qualitatively his scenario seems to predict
certain key traits of UGC~10043: its wind, its dusty inner
disk, and possibly the formation of a distinct inner bulge core,
although quantitatively the agreement is less good.
Typical boundary radii for the shock regions estimated by Wakamatsu
are $\sim4h_{b}$, where $h_{b}$ is the exponential scale length of the
S0 disk (or bulge, in our case). For UGC~10043, exponential fits to
the outer bulge isophotes
yield $h_{b}\approx$1~kpc, suggesting the shock boundary will extend
to roughly 4~kpc. The radial extent of the 
nuclear star-forming region in UGC~10043 was inferred to be $\lsim$2~kpc
(Section~\ref{radio}), 
corresponding rather closely with the observed edge of the bulge, but
roughly a factor of two smaller than predicted by Wakamatsu's model.
Nonetheless, further investigation of this type of model via numerical
modelling would be of interest for exploring the inner gas dynamics of
UGC~10043 and other related galaxies.

Another galaxy that may provide a test of link between Wakamatsu's
scenario and the formation of galactic winds is the polar ring galaxy
NGC~660 (Alton et al. 2000). 
NGC~660 is classified as a ``superwind'' galaxy, and it is also
a relatively rare
example of a polar ring-like structure surrounding a gas-rich disk galaxy, rather
than an early type host, making it a likely environment for the
production of shocks. In addition,
the morphology of the superwind galaxy
NGC~1482 (Veilleux \& Rupke
2002) suggests it too may be a related object.
To our knowledge, no kinematic or structural decoupling has yet been
identified in NGC~1482, but
inspection of 
DSS images reveals 
hints of a ring or extended disk misaligned with the
dust lane of this galaxy. 
Overall it would seem  that 
the search for galactic wind signatures in other polar ring
galaxies or early-type spirals with  kinematically decoupled cores and
gas-rich disks may
yield important clues to the origin of both phenomena.

Finally, we note that if the inner polar ring structure described in
Section~\ref{unsharp} is real, its origin is likely to require an
additional event independent of the formation of the main galaxy. This
would 
most likely have involved the 
accretion of a low-mass, gas-rich satellite, and thus serves as
another possible source for triggering 
the nuclear star formation and wind in UGC~10043. For an orbiting
inner ring,
collisions between gas clouds within the
ring and gaseous material in the disk are expected to produce
compression to drive molecular
gas formation and therefore drive 
star formation (e.g., Tenorio-Tagle
1981; Galletta 1991). 
Hot (shock-heated) gas may also be produced
in this process (Galletta 1991).

\section{Summary\protect\label{summary}}
We have presented new, photometrically-calibrated optical imaging 
($B$, $R$, and \HA) and two-dimensional integral field spectroscopy
of the unusual edge-on spiral galaxy UGC~10043. The 
spectroscopy included measurements of both the stellar \CaII\ infrared
triplet and the \HA\ and [\NII] emission lines from the ionized
gas. Although UGC~10043 has been previously classified as Hubble type Sbc,
our observations have revealed 
a number of structural and kinematic complexities in UGC~10043 that
set it apart from prototypical Sbc spirals.

A non-parametric photometric decomposition of the bulge of UGC~10043
reveals evidence that it is triaxial. Its  inner
isophotes are prolate (elongated perpendicular to the disk),  while
the outer isophotes twist to become oblate and nearly circular. The vertical
elongation and disky isophote shape of the inner bulge region implies that
its triaxiality 
cannot be readily explained by the presence of a bar. Measurements of
the bulge kinematics directly along the midplane ($|z|<4''$) 
are hindered by dust
absorption and contamination from disk light; however at modest
distances from the 
plane ($|z|>5''$), no evidence for systematic rotation  is observed 
in the stellar \CaII\ lines, 
implying the UGC~10043 bulge is slowly
rotating, if at all, along the direction of the disk rotation. 
The UGC~10043 bulge thus appears more similar to an elliptical galaxy
than a typical bulge when placed 
on the $V_{\rm max}/\sigma_{*}$-$\epsilon$ relation.

One hemisphere of the inner bulge of UGC~10043 is girdled by
a narrow dust lane running along the direction of the minor axis. Unsharp
masking has provided tentative evidence that this feature may form part of
an inner polar ring structure roughly 2.5~kpc in diameter. 
The presence of vertically-elongated inner bulge isophotes, encircled
by this dust lane suggests 
a component with misaligned angular momentum in this galaxy. 
However, spectroscopically, we have found no compelling evidence  for
orthogonal rotation, either in the stellar
component or in the ionized gas on scales of $r\gsim5''$. Hence in spite of
exhibiting certain morphological similarities to polar ring and polar disk
galaxies, UGC~10043 is not a kinematically-confirmed 
member of this class. Structurally, UGC~10043  appears
to be closely related to the recently-discovered 
class of spiral galaxies with ``orthogonally decoupled bulges''
(Bertola et al. 1999; Sarzi et al. 2000; Corsini et al. 2003). A
hallmark of such galaxies is the presence of a compact,
orthogonally-rotating inner bulge core; however,
owing to the limited spatial
resolution of our current spectra, coupled with heavy dust
extinction at small radii, we were unable to establish whether
the UGC~10043 bulge may contain an analogous compact ($r\lsim5''$), 
kinematically-decoupled core.  
So far, minor axis dust  lanes have not been reported in other
orthogonally-decoupled bulge systems, and the link between this
structure, a possible decoupled inner core, and the overall formation history of
the bulge is unclear. One possibility is that the minor axis dust
lane formed from the recent accretion of a dwarf satellite and
therefore has an
origin completely independent of the overall bulge structure. Additional, high
spatial resolution spectroscopy is needed to shed
further insight onto the above issue. Unfortunately these observations are
likely to be challenging due to the strong
dust attenuation along the midplane, the requirement for both high
spatial and spectral resolution, and the faintness of
the possible inner polar structure. 

The stellar disk of UGC~10043 also exhibits several interesting properties. It 
is unusually thin for a galaxy of its Hubble
type, with an exponential scale height 
$h_{z}\approx$395~pc. It also has a low optical 
surface brightness, and we estimate a deprojected central surface
brightness $\mu(0)_{R,i}\sim$23.2 mag
arcsec$^{-2}$. While the inner disk regions of UGC~10043 appear to be
extremely dusty, dust in the outer disk appears to be sparse and
patchy, and dust attenuation alone cannot explain the faint appearance
of the disk. The dusty, inner portion of the UGC~10043 disk appears
tilted by $\sim3-5^{\circ}$ relative to the outer disk regions. 
In addition, the outer stellar disk shows a deviation from
planarity in the form of  a weak,
integral sign warp. 

Although \HA, radio continuum, and far-infrared observations all imply
a rather low global star formation rate in UGC~10043 
($\lsim1~M_{\odot}$ yr$^{-1}$),
we have uncovered both photometric and spectroscopic evidence
for a large-scale galactic wind in this galaxy. Narrow-band 
\HA+[\NII] imaging reveals
a biconical distribution of ionized gas
extending to $\sim\pm$3.5~kpc from the plane. The [\NII]
emission lines at most locations in the galaxy resolve into at
least two distinct
velocity components, with a typical separation of 70-80~\kms. The
component closer to the systemic velocity can be ascribed to material
following the disk rotation, while the higher-velocity component
appears to arise from gas outflowing along the surfaces of the
biconical structure. The \HA\ line profiles
also show line splitting 
consistent with the wind model, but in many cases, also show
hints of an additional velocity component close to the systemic
velocity whose origin is unclear.  We estimate an outflow
speed for the UGC~10043 wind of $V_{\rm out}\gsim$104~\kms. This value is
significantly smaller than the outflow speeds of typical ``superwind''
galaxies, and suggests the wind is powered by 
a rather weak central starburst, confined to the inner $r\lsim$2~kpc of
the galaxy. Consistent with other disk galaxies
with large-scale winds, we measure a systematic increase in
the  [\NII]/\HA\
ratio as a function of distance from the midplane in UGC~10043, with
values reaching as large as $\sim$1.7. Shock-heating may be required
to explain these line ratios, although additional line diagnostics and
modelling will be needed to evaluate their importance compared with
photoionization.

The structural and kinematic properties of UGC~10043 seem to be most readily
explained by a formation history involving a ``second event'' which
occurred at least a few Gyr ago.
We have discussed three possible formation mechanisms for UGC~10043,
respectively involving a major
accretion event, a capture, or a major merger. Qualitatively,
each scenario seems able to produce certain features of the UGC~10043
system. However, each model also appears to have serious shortcomings, and
accounting for the thin, gas-rich disk and peculiar bulge structure
of UGC~10043 poses significant challenges. 
Further,
multi-wavelength observations of UGC~10043 together with detailed
numerical modelling are sorely needed in order to supply further
constraints on the evolutionary history of this interesting galaxy.

\acknowledgements 
This work benefited from invaluable contributions from a
number of sources. We are grateful to the UW-Madison TAC for making the
WIYN observing time available, and to the WIYN staff at Kitt Peak for
their observing support.
LDM is indebted to M. Verheijen for help with the reduction of the
DensePak data, and owes thanks to J. Gallagher and L. Smith 
for participation in the WIYN observing, R. Kurucz for making his
synthetic Arcturus spectra available, and H.-W. Chen for
assistance with the Bruzual \& Charlot models. We also acknowledge useful
discussions with a number of other colleagues, including 
N. Caldwell, 
L. Kewley, J. Raymond, P. Schechter, K. Wood, and the attendees of the 
UW-Madison ISM lunch group. Some of the figures for
this paper made use of an IDL script designed by
B. Otte. Finally, we thank J. Gallagher and
L. Sparke for comments on an early draft and the referee, S. Odewahn,
for his suggestions.
Financial support for this work was provided by a Clay Fellowship from the
Harvard-Smithsonian Center for Astrophysics and NASA grant GO-08645.13-A.

\suppressfloats

\suppressfloats

\suppressfloats

\clearpage

\begin{deluxetable}{llc}
\tablewidth{30pc}
\tablenum{1}
\tablecaption{UGC~10043 Basic Data}
\tablehead{\colhead{} & \colhead{} & \colhead{Ref.}}

\startdata

$\alpha$ (J2000.0): &  15 48 41.2  & 1\\
$\delta$ (J2000.0): &  +21 52 09.8 & 1\\
Hubble type: &  Sbc & 1 \\
$A_{B}$: & 0.247 mag & 2 \\
Distance:$^{a}$ &  33.4~Mpc & 3\\

\cutinhead{Measured Quantities}
$D_{R,25.5}$ & \am{2}{8}$\pm$\am{0}{1} & 4\\
PA & \ad{151}{5}$\pm$\ad{0}{5} & 4 \\
$i$ & $90^{\circ}$ & 4\\
$h_{z,R}$$^{b}$ & \as{2}{44}$\pm$\as{0}{04} & 4 \\
$h_{r,R}$$^{c}$ & $40''\pm 6''$ & 5 \\
\\
$h_{r,I}$$^{d}$ & $20''\pm 6''$ & 6 \\
$m_{B}$$^{e}$ & 14.73$^{+0.06}_{-0.06}$ & 4 \\
$m_{R}$$^{e}$ & 13.45$^{+0.10}_{-0.11}$ & 4 \\
$(B-R)$$^{e}$ & 1.28$\pm$0.12 & 4 \\
$m_{K_s}$$^{e}$ & 10.39$\pm$0.03 & 1 \\
$(J-K_{s})$$^{e}$ &  1.32 & 1 \\

\\
$V_{\rm sys}(\CaII)^{f}$ & 2155$\pm$10 km s$^{-1}$ & 4 \\
$V_{\rm sys}({\rm H}\alpha)^{f}$ & 2157$\pm$5 km s$^{-1}$ & 4 \\
$V_{\rm sys}(\NII)^{f}$ & 2161$\pm$6 km s$^{-1}$ & 4 \\

\cutinhead{Derived Quantities}

$M_{B}$$^{g}$ & $-19.4$ & 4 \\
$M_{R}$$^{g}$ & $-20.2$ & 4 \\
$L_{B}$$^{g}$ & $9.0 \times10^{9}~L_{\odot}$ & 4 \\
$M_{R,{\rm bulge}}$$^{h}$ & $-18.5$ & 4 \\
$M_{R,{\rm disk}}$$^{g}$ & $-19.9$ & 4 \\
B/D & 0.3 & 4 \\
$\mu_{R,{\rm obs}}(0)$$^{i}$ & 21.2 mag arcsec$^{-2}$ & 4 \\
$\mu_{R,i}(0)$$^{j}$ & $\sim$23.2 mag arcsec$^{-2}$ & 4 \\
$A_{R,25.5}$ & 27.2~kpc & 4 \\
$L_{{\rm H}\alpha}$$^{k}$ & $\approx1.7\pm0.3\times10^{40}$ erg s$^{-1}$ & 4\\

\cutinhead{Radio and Infrared Properties}

$V_{{\rm sys},\HI}$ & 2161$\pm$1 km s$^{-1}$ & 7 \\
$W_{20}$ & 341$\pm$6 km s$^{-1}$ & 8 \\
$W_{50}$ & 318$\pm$2 km s$^{-1}$ & 7 \\
$V_{\rm max}$$^{l}$ & $\sim$160 km s$^{-1}$ & 4 \\
$M_{\rm dyn}$$^{m}$ & $1.0\times10^{11}~M_{\odot}$ & 4 \\
$\int S_{HI}{\rm d}\nu$ & 22.29 Jy km s$^{-1}$ & 7 \\
$M_{\HI}$ & 5.9$\times10^{9}~M_{\odot}$ & 7\\
$M_{\HI}/L_{B}$ & 0.66 & 4,7 \\
$F_{\rm cont}$ (6-cm) & 4.5$\pm$0.1 mJy & 9 \\
$F_{\rm cont}$ (20-cm) & 10.6$\pm$0.9 mJy & 10 \\
$S_{60\mu m}/S_{100\mu m}$ & 0.34 & 1 \\
FIR$^{l}$ & $(8.1\pm0.7)\times10^{-14}$ W m$^{-2}$ & 1,4 \\ 
$L_{\rm FIR}$$^{m}$ & 2.8$\times10^{9}~L_{\odot}$ & 1,4 \\

\enddata

\tablecomments{Units of right ascension are hours, minutes, and
seconds, and units of declination are degrees, arcminutes, and
arcseconds.}

\tablenotetext{a}{Based on the recessional velocity and Virgocentric
infall correction from the LEDA database and
$H_{0}=$70 km s$^{-1}$ Mpc$^{-1}$.}

\tablenotetext{b}{Average based on exponential fits to 15-pixel-wide
strips perpendicular to the disk at $r$=+45$''$ and $r=-34''$
(locations selected based on minimal dust obscuration; see text).}

\tablenotetext{c}{$R$-band value derived from two-dimensional
fit (exponential in $r$), assuming a
symmetric disk cut-off radius $r_{\rm cut}$=\as{82}{5}.}

\tablenotetext{d}{$I$-band value derived from a one-dimensional
exponential fit between 
$r$=15$''$-40$''$ and assuming no disk cut-off.}

\tablenotetext{e}{Corrected for  Galactic foreground extinction.}

\tablenotetext{f}{As measured in fiber 43.} 

\tablenotetext{g}{Corrected for foreground and internal extinction
(see Section~3.1).}

\tablenotetext{h}{Corrected for Galactic foreground extinction;
internal extinction in bulge assumed to be negligible.}

\tablenotetext{i}{Extrapolated disk central surface brightness,
uncorrected for inclination or internal extinction.}

\tablenotetext{j}{Corrected for foreground and internal extinction and
deprojected to a face-on value following Matthews et al. 1999.}

\tablenotetext{k}{See Section~2.}

\tablenotetext{l}{Assuming $V_{\rm max}\approx\frac{1}{2}(W_{20} -
W_{\rm rand})$, where we adopt 
$W_{\rm rand}$=20~km s$^{-1}$ to account for the random
component of the gas motions.}

\tablenotetext{m}{Total dynamical mass computed from
$M_{\rm dyn}=2.326\times10^{5}rV^{2}_{\rm max}$, where we have taken
$r=1.25R_{25.5}$=17~kpc and $V_{\rm max}$=160~km s$^{-1}$.}

\tablenotetext{n}{Far-infrared flux between 40-120$\mu$m, 
based on the {\it IRAS} 60 and 100$\mu$m fluxes from
the NED database and the definition $FIR=1.26\times10^{-14}(2.58S_{60}
+ S_{100})$ W m$^{-2}$ (Helou et al. 1988). }

\tablenotetext{o}{Far-infrared luminosity, calculated from
$L_{\rm FIR}=4\pi ({\rm FIR})D^{2}/(3.86\times10^{36} {\rm W} L^{-1}_{\odot}$).}

\tablerefs{(1) NED database; (2) Schlegel et al. 1998;  (3) LEDA database;
(4) this work; (5) Pohlen 2001; (6) de Grijs \& van der Kruit 1996;
(7) Giovanelli et al. 1997; (8) Lewis et al. 1985; (9) Hummel et al. 1991
(10) Condon et
al. 1998.}

\end{deluxetable}

\suppressfloats

\begin{deluxetable}{ccccc}
\tablewidth{0pc}
\tablenum{2}
\tablecaption{Photometric Solution}
\tablehead{ \colhead{Filter or Color} & \colhead{$a_{0}$} &
\colhead{$a_{1}$} & \colhead{$a_{2}$} & \colhead{$\sigma_{\rm mag}$}}

\startdata

$B$ &  -0.462 & 0.238 & 0.037 & 0.030 \\

$R$ &   -0.134 &  0.117 & 0.000 &  0.013 \\

$B-R$ & -0.316 & 0.161 & 0.029 & 0.020 \\

\enddata

\tablecomments{A total of 18 standard stars were observed in both $B$ and $R$.
$\sigma_{\rm mag}$ is the RMS dispersion in the fit to the equation:
$m_{\lambda} = 26.0 -2.5~{\rm log(counts/second)} + a_{0} +
a_{1}(X-1) + a_{2}(b-r)$ where $a_{0}$ is the zero point, $a_{1}$ is
the extinction coefficient, $a_{2}$ is the color term, $b-r$ is the
instrumental $B-R$ color, and $X$ is the
airmass.}

\end{deluxetable}

\suppressfloats
\clearpage

\suppressfloats

\suppressfloats

\suppressfloats

\suppressfloats

\suppressfloats

\suppressfloats

\suppressfloats

\suppressfloats

\suppressfloats

\suppressfloats

\suppressfloats

%

\begin{references}
%
Alton, P.B., Stockdale, D. P., Scarrott, S. M., \& Wolstencroft,
R. D. 2000, A\&A, 357, 443

Appleton, P. N. \& Struck-Marcell, C. 1996, Fund. Cosmic Phys., 16, 111

Arnaboldi, M., Capaccioli, M., Cappellaro, E., Held, E. V., \& Sparke,
L. 1993, A\&A, 267, 21

Arnaboldi, M., Capaccioli, M., \& Combes, F. 1994, in The First
Stromlo Symposium: The Physics of Active Galaxies, ASP Conf. Series,
Vol. 54, ed. G. V. Bicknell, M. A. Dopita, and P. J. Quinn, (ASP: San
Francisco), 437

Balcells, M. \& Peletier, R. F. 1994, AJ, 107, 135

Barden, S. C., Sawyer, D. G., \& Honeycutt, R. K. 1998, in 
Optical Astronomical Instrumentation, ed. S. D'Odorico,
Proc. SPIE 3355, 892

Bekki, K. 1998, ApJ, 499, 635

Bell, E. F. 2002, ApJ, 581, 1013

Bender, R., D\"obereiner, S., \& M\"ollenhoff, C. 1988, A\&AS, 74, 385

Bertola, F. 1987, in Structure and Dynamics of Elliptical Galaxies,
IAU Symposium 127, ed. P. T. de Zeeuw (Dordrecht: Reidel), 135

Bertola, F., Buson, L. M., \& Zeilinger, W. W. 1992, ApJ, 401, L79

Bertola, F. \& Corsini, E. M. 2000, in Dynamics of Disk Galaxies: from
the Early Universe to the Present, ASP Conf. Series, Vol. 197,
ed. F. Combes, G. A. Mamon, and V. Charmandaris, (ASP: San Francisco), 115

Bertola, F., Corsini, E. M., Vega Beltr\'an, J. C., Pizzella, A.,
Sarzi, M., Cappellari, M., \& Funes, J. G. S.J. 1999, ApJ, L127

Bertola, F., Vietri, M., \& Zeilinger, W. W. 1991, ApJ, 374, L13

Bettoni, D. \& Galletta, G. 1991, in Dynamics of Disc Galaxies,
ed. B. Sundelius, (Chalmers University: G\"oteborg), 317

Binney, J. \& May, A. 1986, MNRAS, 218, 743

Biretta, J. A. et al. 2000, WFPC2 Instrument Handbook, Version 5.0
(Baltimore: STScI)

Bizyaev, D. \& Mitronova, S. 2002, A\&A, 389, 795

Bland-Hawthorn, J. \& Cohen, M. 2003, ApJ, 582, 246

Bland-Hawthorn, J., Freeman, K. C., \& Quinn, P. J. 1997, ApJ, 490, 143

Bournaud, F. \& Combes, F. 2003, A\&A, 401, 817

Bressan, A., Silva, L., \& Granato, G. L. 2002, A\&A, 392, 377

Bruzual, G. \& Charlot, S. 2003, MNRAS, 344, 1000

Caldwell, N. \& Rose, J. A. 1997, AJ, 113, 492

Carignan, C., C\^ot\'e, S., Freeman, K. C., \& Quinn, P. J. 1997, AJ,
113, 1585

Chen, B. et al. 2001, ApJ, 553, 184

Chevalier, R. A. \& Clegg, A. W. 1985, Nature, 317, 44

Combes, F., Debbasch, R., Friedli, D., \& Pfenniger, D. 1990, A\&A,
233, 82

Condon, J. J. 1992, ARA\&A, 30, 575

Condon, J. J., Cotton, W. D., Greisen, E. W., Yin, Q. F., Perley,
R. A., Taylor, G. B., \& Broderick, J. J. 1998, AJ, 115, 1693

Corsini, E. M., Pizzella, A., \& Bertola, F. 2002, A\&A, 382, 488

Corsini, E. M., Pizzella, A., Coccato, L., \& Bertola, F. 2003, A\&A,
408, 873

Dalcanton, J. J. \& Bernstein, R. A. 2002, AJ, 124, 1328

Davies, R. L., Efstathiou, G., Fall, S. M., Illingworth, G., \&
Schechter, P. L. 1983, ApJ, 266, 41

Debattista, V. P. \& Sellwood, J. A. 1999, ApJ, 513, L107

de Grijs, R. 1998, MNRAS, 299, 595

de Grijs, R. \& Peletier, R. F. 2000, MNRAS, 313, 800

de Grijs, R., Peletier, R. F., \& van der Kruit, P. C. 1997, A\&A, 327, 966

de Grijs, R. \& van der Kruit, P. C. 1996, A\&AS, 117, 19

de Jong, R. S. 1996a, 313, 45

de Jong, R. S. 1996b, A\&A, 313, 377

de Vaucouleurs, G., de Vaucouleurs, A., Corwin, J. G. Jr., Buta,
R. J., Paturel, G., \& Fouqu\'e, P. 1991, Third Reference Catalogue of
Bright Galaxies, Ver. 3.9, (New York: Springer-Verlag)

Eckart, A. \& Downes, D. 2001, ApJ, 551, 730

Elmegreen, B. G. 1999, ApJ, 517, 103

Erwin, P. \& Sparke, L. S. 2003, ApJS, 146, 299

Firmani, C., Hernandez, X., \& Gallagher, J. 1996, A\&A, 308, 403

Freeman, K. \& Bland-Hawthorn, J. 2002, ARA\&A, 40, 487

Freudling, W., Haynes, M. P., \& Giovanelli, R. 1988, AJ, 96, 1791

Gallagher, J. S., Sparke, L. S., Matthews, L. D., Frattare, L. M.,
English, J., Kenny, A. L., Iodice, E., \& Arnaboldi, M. 2002, ApJ,
568, 199

Galletta, G. 1991, in Warped Disks and Inclined Rings around Galaxies,
ed. S. Casertano, P. Sackett, and F. Briggs, (Cambridge: Cambridge
University Press), 207

Garc\'\i a-Ruiz, I., Sancisi, R., \& Kuijken, K. 2002, A\&A, 394, 769

Gerber, R. A., Lamb, S. A., \& Balsara, D. S. 1996, MNRAS, 278, 345

Gerritsen, J. P. E. \& de Blok, W. J. G. 1999, A\&A, 342, 655

Giovanelli, R., Avera, E., \& Karachentsev, I. D. 1997, AJ, 114, 122

Hawarden, T. G., Elson, R. A. W., Longmore, A. J., Tritton, S. B., \&
Corwin, H. G. Jr. 1981, MNRAS, 196, 747

Heckman, T. M. 2001, in Gas and Galaxy Evolution, ASP Conference
Series, Vol. 240, ed. J. E. Hibbard, M. P. Rupen, and J. H. van
Gorkom, (ASP: San Francisco), 345

Heckman, T. M., Lehnert, M. D., Strickland, D. K., \& Armus, L. 2000,
ApJS, 129, 493

Heckman, T. M., Armus, L., \& Miley, G. K. 1990, ApJS, 74, 833

Helou, G., Khan, I. R., Malek, L. \& Boehmer, L. 1988, ApJS, 68, 151

Howk, J. C. \& Savage, B. D. 1999, AJ, 117, 2077

Huang, S. \& Carlberg, R. G. 1997, ApJ, 480, 503

Hummel, E., Beck, R., \& Dettmar, R.-J. 1991, A\&AS, 87, 309

Iodice, E., Arnaboldi, M., De Lucia, G., Gallagher, J. S. III, Sparke,
L. S., \& Freeman, K. C. 2002, AJ, 123, 195

Jergen, J., Binggeli, B., \& Freeman, K. C. 2000, AJ, 119, 593

Just, A., Fuchs, B., \& Wielen, R. 1996, A\&A, 309, 715

Kannappan, S. J., Jansen, R. A., \& Barton, E. J. 2004, AJ, 127, 1371

Karachentsev, I. D., Karachentseva, V. E., Parnovsky, S. L. 1993,
Astron. Nachr., 314, 97

Karataeva, G. M., Yakovleva, V. A., Hagen-Thorn, V. A., \&
Mikolaichuk, O. V. 2001, Ast. Let., 27, 74

Kennicutt, R. C. 1998, ARA\&A, 36, 189

Kewley, L. J., Geller, M. J., Jansen, R. A., \& Dopita, M. A. 2002,
AJ, 124, 3135

Kormendy, J. 1993, in Galactic Bulges, IAU Symp. 153, ed. H. Dejonghe and
H. J. Habing (Dordrecht: Kluwer), 209

Kormendy, J. \& Illingworth, G. 1982, ApJ, 256, 460

Kormendy, J. \& Illingworth, G. 1983, ApJ, 265, 632

Kurucz, R. L. 2003, http://kurucz.harvard.edu/stars.html

Landolt, A. U. 1992, AJ, 104, 340

Lehnert, M. D. \& Heckman, T. M. 1996, ApJ, 462, 651

Lewis, B. M., Helou, G., \& Salpeter, E. E. 1985, ApJS, 59, 161

Liszt, H. S. \& Burton, W. B. 1980, ApJ, 236, 779

Malin, D. F., Quinn, P. J., \& Graham, J. A. 1983, ApJ, 272, L5

Martin, C. L. 1998, ApJ, 506, 222

Matthews, L. D. 2000, AJ, 120, 1764

Matthews, L. D. \& Gallagher, J. S. III. 1997, AJ, 114, 1899

Matthews, L. D. \& van Driel, W. 2000, A\&AS, 143, 421

Matthews, L. D., van Driel, W., \& Monnier Ragaigne, D. 2001, A\&A,
365, 1

Matthews, L. D. \& Wood, K. 2001, ApJ, 548, 150

Merrifield, M. R. 1996, in Barred Galaixes, IAU Colloquium 157, ASP
Conf. Series, Vol, 91, ed. R. Buta, D. A. Crocker, and
B. G. Elmegreen, (San Francisco: ASP), 179

Mihalas, D. \& Binney, J. 1981, Galactic Astronomy, (New York:
W. H. Freeman and Company)

Miller, S. T. \& Veilleux, S. 2003, ApJ, 592, 79

Moriondo, G., Giovanardi, C., \& Hunt, L. K. 1998, A\&AS, 130, 81

Nieto, J.-L. \& Prugniel, P. 1987, A\&A, 186, 30

Oosterloo, T. A., Morganti, R., Sadler, E. M., Vergani, D., \&
Caldwell, N. 2002, AJ, 123, 729

Osterbrock, D. E., Fulbright, J. P., \& Bida, T. A. 1997, PASP, 109,
614

Ostriker, E. C. \& Binney, J. J. 1989, MNRAS, 237, 785

Otte, B., Reynolds, R. J., \& Gallagher, J. S. III. 2001, ApJ, 560, 207

Pohlen, M. 2001, PhD Thesis, Ruhr-Universit\"at Bochum

Pohlen, M., Dettmar, R.-J., L\"utticke, R., \& Schwarzkopf, U. 2000,
A\&AS, 144, 405

Pitesky, J. 1991, in Warped Disks and Inclined Rings around Galaxies,
ed. S. Casertano, P. Sackett, and F. Briggs, (Cambridge: Cambridge
University Press), 215

Pizzella, A., Corsini, E. M., Morelli, L., Sarzi, M., Scarlata, C.,
Stiavelli, M., \& Bertola, F. 2002, ApJ, 573, 131

Rand, R. J. 1998, ApJ, 501, 137

Reshetnikov, V. \& Combes, F. 1997, A\&A, 324, 80

Reshetnikov, V. \& Combes, F. 1998, A\&AS, 138, 101

Reshetnikov, V. P., F\'aundez-Abans, M., \& de Oliveira-Abans, M. 2002,
A\&A, 383, 390

Reshetnikov, V. P., Hagen-Thorn, V. A., \& Yakovleva, V. A. 1995,
A\&A, 303, 398

Reshetnikov, V. \& Sotnikova, N. 1997, A\&A, 325, 933

Rix, H.-W., Franx, M., Fisher, D., \& Illingworth, G. 1992, ApJ, 400,
L5

Rix, H.-W. \& Katz, N. 1991, in Warped Disks and Inclined Rings around Galaxies,
ed. S. Casertano, P. Sackett, and F. Briggs, (Cambridge: Cambridge
University Press), 112

Roberts, M. S. \& Haynes, M. P. 1994, ARA\&A, 32, 115

Sackett, P. D. \& Sparke, L. S. 1990, ApJ, 361, 408

Sage, L. J. \& Galletta, G. 1993, ApJ, 419, 544

Sandage, A. \& Bedke, J. 1994, The Carnegie Atlas of Galaxies,
(Washington: Carnegie Institution of Washington)

Sanders, D. B., Solomon, P. M., \& Scoville, N. Z. 1984, 276, 182

Sarzi, M., Corsini, E. M., Pizzella, A., Vega Beltr\'an, J. C.,
Cappellari, M., Funes, J. G. S. J., \& Bertola, F. 2000, A\&A,  360,
439

Schlegel, D. J., Finkbeiner, D. P., \& Davis, M. 1998, ApJ, 500, 525

Schweizer, F., Whimore, B. C., \& Rubin, V. C. 1983, AJ, 88, 909

Siegel, M. H., Majewski, S. R., Reid, I. N., \& Thompson, I. B. 2002,
ApJ, 578, 151

Sil'chenko, O. K. 2002, Ast. Let., 28, 207

Sparke, L. S. 1986, MNRAS, 219, 657

Sparke, L. S. 1996, ApJ, 473, 810

Sparke, L. S. 2002, in Disks of Galaxies: Kinematics, Dynamics, and
Perturbations, ASP Conf. Series, Vol. 275, ed. 
E. Athanassoula, A. Bosma, and R. Mujica (ASP: San Francisco), 367

Suchkov, A. A., Balsara, D. S., Heckman, T. M., \& Leitherner,
C. 1994, ApJ, 430, 511 

Tenorio-Tagle, G. 1981, A\&A, 94, 338

Terlevich, E., D\'\i az, A. I., Terlevich, R., Gonz\'alez-Delgado,
R. M., P\'erez, E., \& Garc\'\i a Vargas, M. L. 1996, MNRAS, 279, 1219

Tohline, J. E., Simonson, G. F., \& Caldwell, N. 1982, ApJ, 252, 92

Tonry, J. \& Davis, M. 1979, AJ, 84, 1511

T\'oth, G. \& Ostriker, J. P. 1992, ApJ, 389, 5

Tully, R. B., Pierce, M. J., Huang, J.-S., Saunders, W., Verheijen,
M. A. W., \& Witchalls, P. L. 1998, AJ, 115, 2264

Tully, R. B. \& Verheijen, M. A. W. 1997, ApJ, 484, 145

Veilluex, S. 2004, in Recycling Intergalactic and Interstellar Matter,
IAU Symp. Series, Vol. 217, ed. P.-A. Duc, J. Braine, and E. Brinks,
in press (astro-ph/0310585)

Veilleux, S. \& Rupke, D. S. 2002, ApJ, 565, 63

Vel\'azquez, H. \& White, S. D. M. 1999, MNRAS, 304, 254

Wainscoat, R. J., Freeman, K. C., \& Hyland, A. R. 1989, ApJ, 337, 163

Wakamatsu, K.-I. 1993, AJ, 105, 1745

Whitmore, B. C., Lucas, R. A., McElroy, D. B., Steiman-Cameron, T. Y.,
Sackett, P. D., \& Olling, R. P. 1990, AJ, 100, 1489

Whitmore, B. C., McElroy, D. B., \& Schweizer, F. 1987, ApJ, 314, 439

Windhorst, R. A., Taylor, V. A., Jansen, R. A., Odewahn, S. C.,
Chiarenza, C. A. T., Conselice, C. J., de Grijs, R., de Jong, R. S.,
MacKenty, J. W., Eskridge, P. B., Frogel, J. A., Gallagher, J. S. III,
Hibbard, J. E., Matthews, L. D., \& O'Connell, R. W. 2002, ApJS, 143,
113

Wirth, A. \& Gallagher, J. S. III. 1984, ApJ, 282, 85

Wood, K. \& Mathis, J. S. 2004, MNRAS, in press

Zasov, A. V., Makarov, D. I., \& Mikhailova, E. A. 1991,
Sov. Astron. Let. 17, 374

Zeilinger, W. W., Galletta, G., \& Madsen, C. 1990, MNRAS, 246, 324

\end{references}
\end{document}